\documentclass{article}

\usepackage{microtype}
\usepackage{graphicx}
\usepackage{subfigure}
\usepackage{booktabs} 

\usepackage{hyperref}
\usepackage{enumitem}

\usepackage[accepted]{icml2026}

\usepackage{amsmath}
\usepackage{amssymb}
\usepackage{mathtools}
\usepackage{amsthm}

\usepackage{listings}
\usepackage{bm}
\usepackage[breakable]{tcolorbox}
\usepackage{pifont}
\usepackage{comment}
\usepackage{url}

\usepackage[capitalize,noabbrev]{cleveref}
\usepackage{natbib}

\theoremstyle{plain}

\theoremstyle{definition}

\theoremstyle{remark}

\usepackage[textsize=tiny]{todonotes}
\usepackage{comment}

\floatname{algorithm}{Algorithm}
\renewcommand{\algorithmicrequire}{\textbf{Input:}}
\renewcommand{\algorithmicensure}{\textbf{Output:}}

\newcommand{\jon}[1]{{\color{blue} Jon: #1}}

\newcommand{\code}[1]{\texttt{#1}}

\icmltitlerunning{SynPAT: Generating Synthetic Physical Theories with Data}

\begin{document}
%

%

\twocolumn[
  \icmltitle{SynPAT: A System for Generating Synthetic Physical Theories with Data}



  \icmlsetsymbol{equal}{*}

  \begin{icmlauthorlist}
    \icmlauthor{Karan Srivastava}{xxx}
    \icmlauthor{Jonathan Lenchner}{yyy}
    \icmlauthor{Joao Goncalves}{yyy}
    \icmlauthor{Mark S.\ Squillante}{yyy}
    \icmlauthor{Lior Horesh}{yyy}
  \end{icmlauthorlist}

  \icmlaffiliation{xxx}{Department of Mathematics, University of Wisconsin-Madison, Madison, WI, USA}
  \icmlaffiliation{yyy}{Mathematics of Computation Department, IBM T.J. Watson Research Center, Yorktown Heights, NY, USA}

  \icmlcorrespondingauthor{Jonathan Lenchner}{lenchner@us.ibm.com}


\vskip 0.25in
]



\printAffiliationsAndNotice{}  

\begin{abstract}
Machine-assisted methods for discovering physical laws from background theory and data have recently emerged, promising to advance our understanding of the physical world. However, training and benchmarking these systems remains challenging: real physical theories are limited in number. To address this need, we introduce SynPAT, a system for generating synthetic physical theories with accompanying data. SynPAT produces: (i) a consistent set of axioms forming a synthetic theory, (ii) a symbolic consequence of these axioms representing the discovery target, and (iii) noisy data approximating this consequence. Crucially, to mirror historically incorrect theories (e.g., Newtonian mechanics before Special Relativity), SynPAT can also generate theories whose axioms do not strictly entail, and in fact conflict with, the observed consequence, requiring a correction to the assumed axioms to bridge the gap. We detail SynPAT's methodology and benchmark several open-source symbolic regression systems on our generated theories and data.
\end{abstract}

\section{Introduction}
Machine-assisted methods for discovering new physical laws of nature, starting from a given background theory and data, have recently emerged, and seem to hold the promise of someday advancing our understanding of the physical world \cite{ai-descartes23, ai-hilbert24}.  
Given a symbolic regression (SR) problem, in other words, some data representing a phenomenon of which one wishes to capture a symbolic representation, two interesting questions emerge: 1) Can a complete and correct background theory be used to improve on the results of the 
SR
task? 2) Can the presence of a background theory that incorrectly predicts the results of the phenomenon for which we have data, in fact, be helpful in the
SR
task? In practice, the presence of the background theory may actually bias us away from finding a good symbolic form.  But this is the problem faced by leading scientists today in the physical sciences, particularly in physics. We have a solid theory that predicts many phenomena very accurately, but fails in some
cases.  Examples include data on supernova explosions and the rotation of galaxies being at odds with the equations of general relativity, and the observed abundance of lithium being at odds with the equations that we believe govern the strong and weak nuclear forces and nuclear synthesis in the primordial universe.
However, the fact that there are relatively few known theories in the physical sciences has made it difficult to train, test, and benchmark systems that utilize the combination of a background theory and data. 

We sought to address these fundamental challenges by developing SynPAT, a system for generating synthetic physical theories with data. While numerous benchmarking systems exist for discovering individual equations (in isolation) from data, we are the first to provide a benchmarking system that incorporates entire physical theories. Our SynPAT system generates: (i) a set of consistent axioms (the synthetic theory); (ii) a symbolic expression that is a consequence of the axioms and, moreover, the challenge to be discovered; and (iii) noisy data that approximately match the consequence. To mirror the real-life case where physical theories are not quite correct (e.g., the case of Newtonian mechanics before the discovery of Special Relativity), we also generate theories that do not correctly predict the consequence. We provide a detailed description of the inner workings of SynPAT and its various capabilities. We also report on our benchmarking of several open-source
SR
systems using our generated theories and data.

The remainder of this paper is organized as follows. In the next section, we review related work on 
SR,
the benchmarking of 
SR
systems, and recent work employing theories in conjunction with data to discover new scientific laws. 
Next,
we present a detailed description of our theory, consequence, and data generation processes. We then discuss the case where some portion of a theory is not such a good match to the data and the generation of what we call theories with biased data.
Next,
we review the benchmarks we have run using our theories, discuss code and data set availability, and end with conclusions and future work.
Additional technical details and results are presented in the appendices.

\section{Related Work}

SR refers to the idea that, given some data associated with a set of variables, one can search through a space of mathematical expressions to come up with a symbolic expression from the search space that best characterizes the data in the sense of minimizing a given error metric. This idea in its full generality dates back to 1992 and Chapter 2.4 of the textbook on genetic programming
\cite{koza1992}. 
However, the BACON.5 system of 
Langley et al.~\yrcite{langley1981bacon},
which discovered physical conservation laws from a small set of data-driven heuristics, dates back even further -- to 1981. Linear and multilinear regression date back much earlier with the discovery of linear regression actually being credited to Newton \cite{GROPING16}.  

\subsection{Data-driven Approaches}

Some of the early work on 
SR
includes \cite{korns06}, \cite{duffy99}, and \cite{bautu95} for discovering individual equations not involving derivatives, and then the work of 
Bongard \& Lipson~\yrcite{bongard-pnas07}, 
which allows not only for the discovery of individual equations with ordinary derivatives among the regressed symbols but \emph{coupled systems} of such equations. 
Given some data and the problem of finding one or more best-fitting governing ordinary differential equations (ODEs), with initial conditions either provided or output as part of the solution, several recent approaches have utilized deep neural networks (NNs) and even transformers. There is a rapidly growing literature, but some examples include NODE \cite{NODE2018}, ODEFormer \cite{d'ascoli2024odeformer}, which comes with a dataset of 
ODEs from the literature that the authors refer to as ODEBench, and the predecessor system to ODEFormer \cite{10.5555/3618408.3618491}. Both ODEFormer \cite{d'ascoli2024odeformer} and its predecessor system borrow ideas from
La Cava et al.~\yrcite{cava2018learning}, 
which originated the concept of learning concise representations by evolving networks of trees of operators. A nice comparison of some of the most important 
SR
methods and systems up through 2021 is given in \cite{la2021contemporary}, which also includes a large benchmark and benchmark dataset (to be discussed later). 

At least two commercially available
SR
systems exist: Eureqa (software that is now part of the DataRobot software suite)
, whose underlying algorithm is described in \cite{doi:10.1126/science.1165893}, and the Wolfram SR system,  described in \cite{wolfram2004}. Both Eureqa and Wolfram use genetic algorithms. The Eureqa system was judged to be state of the art by the AI Feynman system \cite{doi:10.1126/sciadv.aay2631, udrescu2020aifeynmanphysicsinspiredmethod}, which is one of the five systems we benchmark. 

The AI Feynman system is a ``physics-inspired'' 
SR
system. It recursively applies a set of decomposition tactics and specialized solvers gleaned from the study of known physics equations, followed by polynomial fitting and brute-force search. If the problem is not directly solved, it appeals to a deep 
NN-based
component. As usual with most 
SR
systems, AI Feynman starts with data and tries to discover a single best governing \emph{function}, given user-specified variables and constants, each of which has associated units of measure. It does not support derivatives. 

The AI Feynman system was able to recover all 100 equations it collated from the Feynman Lectures on Physics, Volumes I--III \cite{feynman1963}, using a feedforward
NN
with 6~hidden layers and 100,000 synthetically generated exact data points per equation. The results were compared to a 71\% success rate on the same test set for the Eureqa SR system. As acknowledged by the authors, there is an inherent mixing of training and test data in these results, so they treat these 100~equations as training data for a further test on what they call 20~``bonus'' equations culled from a variety of additional classical physics textbooks. The AI Feynman system was not further tweaked for these additional equations, but again, in response to 100,000 exact synthetically generated data points per equation, the system was able to recover 90\% of the equations versus only 15\% when running Eureqa. 

Following the tests on exact data, the AI Feynman system was also run on data generated with systematic multiplicative Gaussian noise applied to the function values starting at a multiplicative noise level $\epsilon = 10^{-6}$ and increasing by powers of 10 until each equation could no longer be rediscovered. Noisy data results were presented only for the $100$ Feynman Lecture series problems. The minimum value of $\epsilon$ for which each problem is still solvable was reported, with the authors remarking that ``most of the equations can still be recovered exactly with an $\epsilon$-value of $10^{-4}$ or less, while almost half of them are still solved for $\epsilon = 10^{-2}$''.

A second system we benchmark in this work is PySR \cite{pysr, tonda25}. Unlike AI Feynman's physics-inspired approach which uses heuristics such as dimensionality analysis, compositionality and symmetry detection, PySR is designed to be a generalized, physics-agnostic, 
SR
model that uses genetic programming and maintains diverse populations of equations that evolve in parallel. 

A third system we benchmark is the Bayesian Machine Scientist \cite{doi:10.1126/sciadv.aav6971}, which uses a Markov Chain Monte Carlo 
based algorithm to address the goodness of fit versus model complexity tradeoff encountered by all SR systems, learning prior plausibility of models by analyzing 4080 mathematical expressions gleaned from Wikipedia.

A fourth system we include in our benchmarking is the genetic programming-based GPG system, available at \cite{gpg}, which is a reimplementation of the GP-GOMEA (Gene-pool Optimal Mixing Evolutionary Algorithm) system described in \cite{10.1162/evco_a_00278}.

\subsection{Data-and-Theory Driven Approaches}\label{subsec:data-and-theory}

The final 
SR
package we benchmark in this work is the recent system known as AI Hilbert \cite{ai-hilbert24}. Both AI Hilbert and its predecessor system, AI Descartes \cite{ai-descartes23}, solve a fundamentally different problem from that of the other 
SR
systems we have considered thus far; namely, they consider the case where one is given a physical (background) theory represented by a set of equations, or what they call \textit{axioms}, together with some potentially noisy data in a subset of the same variables as the theory. They then attempt to find a consequence of the theory that (i) is a good fit to the data and (ii) is as faithful to the theory as possible, while (iii) favoring simplicity of expression of the generated consequence.
In AI Hilbert, the theories and generated consequences are limited to equations that are multivariate polynomials in the variables and constants provided. So, e.g., exponentials and trigonometric functions are not supported and neither are derivatives. Unlike other SR systems, AI Hilbert and its predecessor system make use of non-linear optimization methods along with theorem proving machinery. When a new consequence of a given theory is deducible from the background theory, a derivation (i.e., proof of derivability of the consequence from the axioms) is also generated.

An even more recent addition to the literature of systems that incorporate background theories is AI Noether \cite{ai-noether25}. 
However, AI Noether does not use data and addresses the problem of a gap in the background theory. Given an equation expressed as a polynomial in some set of variables that is \emph{not} derivable from (but consistent with) a background theory, the AI Noether system tries to arrive at a ``simplest'' additional axiom to allow derivation of the equation. Although the focus of our work is fundamentally a regression task (in other words, explaining data via an equation in some set of unknowns), our benchmarking system can readily be adapted to the AI Noether use case: we could simply generate a synthetic axiom system along with a consequence, and then drop one of the axioms and ask AI Noether to discover the discarded axiom. 

\subsection{Benchmarking Data Sets}

Several 
SR
benchmarking data sets comprising varying numbers of individual equations and accompanying data have been compiled over the course of time.  We have already mentioned the 100 equations from the Feynman lectures and 20 ``bonus'' equations from leading physics text books, along with 100,000 synthetically generated data points, as mentioned in \cite{udrescu2020aifeynmanphysicsinspiredmethod} and available at \cite{f-srdb2020}. A small set of non-linear first-degree 
ODEs
gleaned from Strogatz's textbook on nonlinear dynamics and chaos \cite{strogatz2000}, together with accompanying data, has been compiled in the so-called ODE-Strogatz repository \cite{stro-repo2020}. 
Additional compilations include \cite{egklitz2020BenchmarkingSS}, which includes several synthetic equations as well as equations from a variety of engineering disciplines, and \cite{10.1145/3205455.3205539}, which pulled 94 real-world regression equations from the Penn Machine Learning Benchmark (PMLB) \cite{PMLB17}. Finally, it is important to mention the compilation known as SRBench \cite{la2021contemporary}, which assembled 252 SR regression equations for use in a large benchmark of SR and machine learning methods. All the equations from this study were imported into the previously mentioned Penn Machine Learning Benchmark (PMLB). 
We believe we are the first to provide benchmarks that combine not only single equations with data but entire physical theories with data. 
Having this latter type of benchmark is important to see whether systems that incorporate entire or partial theories in addition to data, such as AI Descartes \cite{ai-descartes23} and AI Hilbert \cite{ai-hilbert24}, perform noticeably better than systems that appeal to data alone for various regression tasks.

\section{Theory, Consequence, and Data 
Generation}

\begin{figure}[ht]
    \centering
    \includegraphics[scale=0.2]{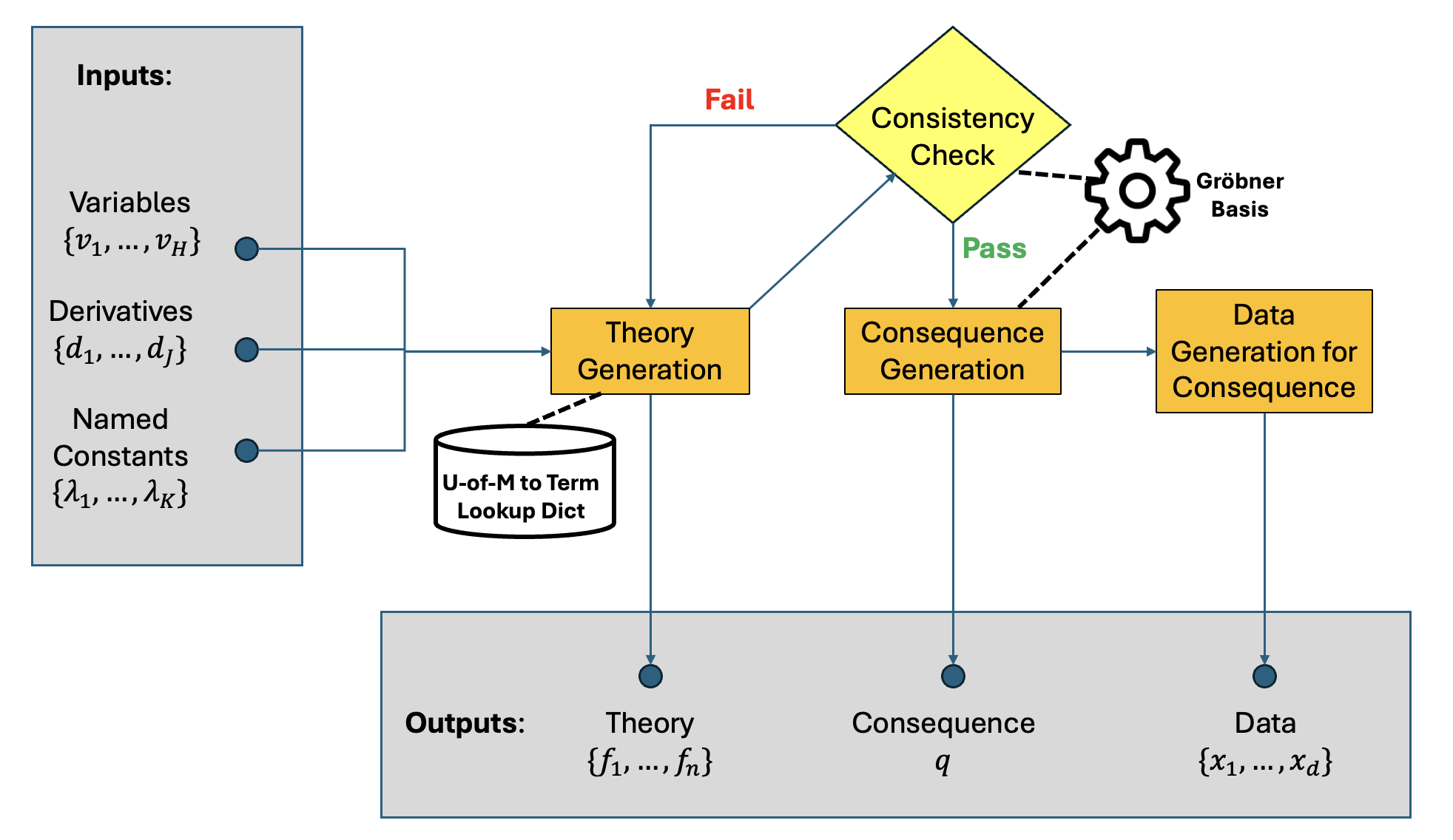}
    \caption{The pipeline for generating a theory, then a consequence, and then data for the consequence. }
    \label{fig:axiom-conseqeunce-pipeline}
\end{figure}

In this section, we describe, in turn, theory generation, consequence generation (a process that includes theory consistency checking), and data generation. A high-level schematic of this pipeline is given in Figure \ref{fig:axiom-conseqeunce-pipeline}. 

\subsection{Theory Generation}\label{sec:Theory-Generation}  
We refer to the basic equations of our theories as ``axioms''.  
Our synthetic theory generation system, SynPAT, starts with a fixed set of variables $\mathbf{V} = \{v_1,...,v_H\}$, ordinary derivatives $\boldsymbol{\Delta} = \{d_1,...,d_J\}$, and named constants $\boldsymbol{\Lambda} = \{\lambda_1,...,\lambda_K\}$ (physical constants such as the gravitational constant, $G$, or mathematical constants such as $\pi$ and $e$), and then generates a user-specified number of equations $\{f_1,...,f_n\}$ based on $\mathbf{V}$, $\boldsymbol{\Delta}$ and $\boldsymbol{\Lambda}$, trying to use as many of the $H$ variables, $J$ derivatives, and $K$ constants as possible. The variables, derivatives, and named constants can have associated units of measure, in which case it is possible to request of SynPAT that it generate only dimensionally consistent axioms. Moreover, it is possible to start with a given theory and ask the system to replace a given (or randomly selected) axiom with a freshly generated axiom or to add to the existing set an additional axiom. 
In addition to ordinary variables, SynPAT allows a distinguished, \emph{dimensionless} argument variable \(\theta\) and \emph{dimensionless} auxiliary symbols \(\sin\_\theta\), \(\cos\_\theta\), and \(e\_\theta\), which respectively denote \(\sin(\theta)\), \(\cos(\theta)\), and \(e^{\theta}\) at \emph{data-generation time} and are in the set of variables $\{v_1,...,v_H\}$, unless otherwise stated.
The system generates equations with integer powers of the chosen symbols.
Generated equations may be viewed as multivariate polynomial equations with the requested variables, derivatives, and constants regarded as indeterminates. 

SynPAT is built in Python on top of the open-source and BSD-licensed Python library for symbolic computation known as SymPy\footnote{\url{https://www.sympy.org/en/index.html}}. The important base classes in SynPAT manage units of measure (classes  \textbf{BaseUnit} for basic units of measure and \textbf{UnitOfMeasure} for all other (derived) units of measure), named constants, variables, and derivatives. Since there is already a \textbf{Derivative} class in SymPy for handling symbolic differentiation, and the name Differential has a somewhat different operator-theoretic meaning within SymPy (and additionally causes namespace collisions), we have given the derivative class in SynPAT the somewhat unusual name \textbf{Derivatif}. \textbf{Derivatif}s are similar to variables (of class \textbf{Variable}) but in addition to having attributes for the name and unit of measure, they also track the dependent and independent variables associated with the derivative along with the order of the derivative. For example, if the derivative we are modeling is $\frac{d^2x}{dt^2}$, then the Python name of the \textbf{Derivatif} might be $d2xdt2$, with independent variable~$t$, dependent variable~$x$, and order~$2$. The \textbf{Variable} and \textbf{Derivatif} classes inherit from SymPy's \textbf{Symbol} class, while SynPAT's \textbf{Constant} class, used for named constants, inherits from \textbf{Variable} and is simply thought of as a \textbf{Variable} with a non-changing (constant) value. 

The bulk of the theory generation work in SynPAT is performed in its \textbf{Equation} and \textbf{EquationSystem} classes. 
Since there are no special functions in SynPAT, equations are generated as a set of terms in a user-specified set of variables, ordinary derivatives, and named constants. 
Terms can be given arbitrary signs so that, for example, $t_1 + t_2 - t_3 + t_4$ is simply the same as the equation $t_1 + t_2 + t_4 = t_3$, while $t_1 + t_2 + t_3 + t_4$ is the same as $t_1 + t_2 + t_3 + t_ 4 = 0$. Note that an equation with ratios of terms, for example, an equation of the form $\frac{t_1}{t_2} + t_3 = \frac{t_4}{t_5}$, can be cleared of its denominators upon multiplying by the least common multiple of the terms $t_2 $ and $t_5$. Hence WLOG we do not need to consider such fractional expressions. 

A description of the process SynPAT uses to randomly generate terms is given in Appendix~\ref{app:term-gen}, along with pseudocode.  With random term generation in hand, in order to generate random not-necessarily dimensionally consistent equations, a random number of terms must be determined in a principled manner (we are guided by the general form of equations from the AI Feynman dataset \cite{f-srdb2020}), and a variety of other checks performed. Details of this equation generation process are provided in Appendix~\ref{app:eqn-gen-not-dim-consistent}. The generation of a dimensionally consistent equation is a considerably more complex process. The key component of this process is the generation of a dictionary that maps every possible compound unit of measure to the set of all terms which match that unit of measure. A detailed description of how this dictionary is created and then used to create dimensionally consistent equations is provided in Appendix~\ref{app:eqn-gen-dim-consistent}, which also includes pseudocode. 

The process for generating a random dimensionally consistent set of equations is largely similar to that for generating an arbitrary set, except that it calls a version of the equation generator which uses dimensional consistency. We focus on the dimensionally consistent case here, since there are a few additional nuances, such as $\theta$ and trig functions of $\theta$, namely $\sin{\theta}$ and $\cos{\theta}$. We also include the trig identity 
$\sin^2(\theta) + \cos^2(\theta) - 1 = 0$ as part of the axioms when relevant. See Appendix~\ref{app:eqn-sys-gen} for details, again including pesudocode. The system of equations now needs to be checked for consistency~--~specifically, whether it admits any satisfying assignments to its variables and derivatives. While consistency checking is conceptually distinct from consequence generation, the two are closely related, as we will see, and is discussed in the next subsection.


%

\subsection{Consequence Generation and Consistency Checking}\label{Projection exposition}

Once we generate a system of axioms, we borrow from the theory of Gr\"{o}bner bases and commutative algebra \cite{coxOshea} to generate consequences of the theory and check the consistency of the axiom system.
We illustrate this with an example of the given sample system:
\begin{gather}
\underline{\textbf{Axioms}} \notag\\
\frac{d^2x_1}{dt^2} + \frac{dx_1}{dt}w = 2\frac{d^2x_2}{dt^2} - 2\frac{dx_2}{dt}w \label{eg1} \\
m_2w\bigg(F_gd_2 - 2\left(\frac{dx_1}{dt}\right)^2\bigg) = \frac{dx_2}{dt}F_gm_1  \label{eg2} \\
G(m_1 - m_2) + d_1d_2\bigg(\frac{d^2x_1}{dt^2} + d_2w^2\bigg) = 0 \label{eg3} \\
\frac{dx_1}{dt}(d_1 + d_2) = \frac{dx_2}{dt}d_1 \label{eg4} \\
\rule{\linewidth}{0.4pt} \notag \\
\textbf{Consequence: } m_2 = \frac{d_1^2d_2\left(\frac{d^2x_2}{dt^2} + wd_2^2\right)}{G(d_1 + d_2)}  \label{egconseq}
\end{gather}

To derive the consequence from the axioms, we first solve equation (\ref{eg4}) for \( \frac{dx_1}{dt} \) and substitute this into equation (\ref{eg1}) to express \( \frac{d^2x_1}{dt^2} \) in terms of \( \frac{d^2x_2}{dt^2} \) and \( \frac{dx_2}{dt} \). We then substitute \( \frac{d^2x_1}{dt^2} \) into equation (\ref{eg3}) and solve for \( m_1 \) and finally substitute \( m_1 \) and \( \frac{dx_1}{dt} \) into equation (\ref{eg2}) to isolate \( m_2 \). Each derivation step corresponds to an algebraic combination of the axioms, and thus the consequence can, after clearing the denominators, 
alternatively be expressed as \begin{small}$$\frac{d_1}{2} \, \text{eq(\ref{eg1})} - \frac{1}{2} \, \text{eq(\ref{eg2})} + \frac{1}{2} \, \text{eq(\ref{eg3})} \, \cdot \,\text{eq(\ref{eg4})} +\frac{d_1(d_1+d_2)}{2} \, \text{eq(\ref{eg4})} = 0,$$ \end{small} 
where each $\text{eq}(i)$ denotes the $i$-th axiom rewritten in homogeneous form (with all terms moved to the left-hand side and set equal to zero). Notably, if this process leads to deriving a contradiction such as $1=0$, it directly shows that the axiom system is inconsistent. 

We follow and extend the terminology of AI Hilbert and call the subset of variables, derivatives, and named constants in the consequence the set of ``measured variables'', ``measured derivatives'', and ``observed constants'', respectively. Indeed, it is this equation or consequence for which we will generate data and thus, from the perspective of the user or scientist interested in creating a more refined theory, the variables (and derivatives) that are effectively measured and observed. For equation (\ref{egconseq}), the measured variables are $d_1,d_2,w$, the observed constant is $G$, and the measured derivative is $\frac{d^2x_2}{dt^2}$. 

In addition to $\mathbf{V} = \{v_1,...,v_H\}$, $\boldsymbol{\Delta} = \{d_1,...,d_J\}$, and $\boldsymbol{\Lambda} = \{\lambda_1,...,\lambda_K\}$, let $\mathbf{v} = \{v_1,...,v_h\}$, $\boldsymbol{\delta} = \{d_1,...,d_j\}$, and $\boldsymbol{\lambda} = \{\lambda_1,...,\lambda_k\}$ be the sets of measured variables, measured derivatives, and observed constants, respectively, where $h\leq H$, $j\leq J$, $k\leq K$. 
To automate the process of finding a consequence over constants and measured unknowns, we will borrow notation and results from computational algebraic geometry. Define $R = \mathbb{R}[\mathbf{V},\boldsymbol{\Delta},\boldsymbol{\Lambda}]$ to be the set (or ring) of polynomials on the variables, derivatives, and constants as indeterminates with real coefficients. \textcolor{black}{When \(\theta\) and its auxiliaries appear, we include all four symbols in the polynomial ring and treat them as indeterminates during elimination. This decouples transcendental semantics from algebraic elimination; the only semantic constraint enforced at the algebraic level is the trigonometric identity \(\sin\_\theta^2 + \cos\_\theta^2 - 1 = 0\) when both \(\sin\_\theta\) and \(\cos\_\theta\) occur among the axioms.
}Each of the axioms is an element of this set. Given polynomials $f_1,...,f_n \in R$, we can define the \textit{ideal} $\langle f_1,...,f_n\rangle$ as the set of finite algebraic combinations of the $f_i$s, i.e., $I = \{\sum_{i=1}^n\alpha_i f_i: \alpha_i\in R\}$. We call $f_1,...,f_n$ a generating set of $I$. As we have seen, if a consequence is derivable from a list of axioms, it can be expressed as an algebraic combination of those axioms. Therefore, the set $I$ is the set of algebraic consequences of the axiom. It follows that the system $f_1,...,f_n$ is inconsistent if and only if the constant $1$ is in $I$ \cite{coxOshea}, that is to say, the contradiction $1=0$ is derivable from the axioms.

We already observed that performing derivations from a set of axioms can lead to a consequence defined over a reduced set of variables, derivatives, and named constants compared to the original system. Algebraically, we are computing $I \cap \mathbb{R}[\mathbf{v},\boldsymbol{\delta},\boldsymbol{\lambda}]$. 
Although this set is infinite, it turns out to be an ideal itself \cite{coxOshea} with a generating set that can be computed using the theory of Gr\"{o}bner bases. To do this, we define a lexicographic monomial order on \( \mathbb{R}[\mathbf{V}, \boldsymbol{\Delta}, \boldsymbol{\Lambda}] \) by taking the subset \( \mathcal{S} = \mathbf{v} \cup \boldsymbol{\delta} \cup \boldsymbol{\lambda} \subset \mathbf{V} \cup \boldsymbol{\Delta} \cup \boldsymbol{\Lambda} \), placing all indeterminates in \( \mathcal{S} \) before those not in \( \mathcal{S} \), and arbitrarily ordering the elements within \( \mathcal{S} \). The Gr\"{o}bner basis of $I$ with respect to this ordering is a generating set $\mathcal{G}$ of $I$ with the following property (known as the \emph{Elimination theorem} \cite{coxOshea}):
$$I \cap \mathbb{R}[\mathbf{v},\boldsymbol{\delta},\boldsymbol{\lambda}] = \langle \mathcal{G}\text{ } \cap \mathbb{R}[\mathbf{v},\boldsymbol{\delta},\boldsymbol{\lambda}] \rangle. \label{elimThm}$$

Therefore, to check if a system $\{f_1,...,f_n\}$ is consistent and generate a consequence over $\mathbf{v},\boldsymbol{\delta},\boldsymbol{\lambda}$, we define the ideal $I = \langle f_1,...,f_n \rangle$ and compute a Gr\"{o}bner basis $\mathcal{G}$ of $I$  using an existing implementation of Buchberger's algorithm \cite{Buchberger, coxOshea} in the computer algebra software Macaulay2 \cite{M2}.  If \( \mathcal{G} = \{1\} \), indicating inconsistency, we discard the system. If not, we compute the intersection $\mathcal{G} \cap \mathbb{R}[\mathbf{v},\boldsymbol{\delta},\boldsymbol{\lambda}]$ and pick the first polynomial in the basis to be the polynomial derived from the theory. However, for some choices of \( \mathbf{v}, \boldsymbol{\delta}, \boldsymbol{\lambda} \), a derived polynomial may not exist or the resulting consequence may be uninformative, such as a monomial relation like \( d_1 d_2 = 0 \). Moreover, for data generation purposes, as we will see in the following section, we restrict consequences to involve at most one observed constant (a constraint that can be adjusted by the user~--~details noted in Section \ref{CodeAvail}). 

To control downstream numerical stability and dataset complexity, we impose a term-count filter on projected consequences: we only accept a candidate whose leading (first) polynomial has at most \(T\) additive terms (default \(T=8\), user-adjustable). Operationally, we begin by randomly selecting a subset of variables, derivatives, and constants, compute a Gr\"obner basis under the corresponding lexicographic order, and inspect the first element of the eliminated basis. If the basis is trivial, the consequence violates other constraints (e.g., constant-limit), or the first polynomial exceeds the \(T\)-term threshold, we reject the candidate and resample the measured subset (or iteratively expand it by adding more variables, derivatives, or constants) and repeat, up to a fixed attempt budget. To avoid trivialities, we never choose measured subsets that coincide with (or are contained in) the support of any single axiom, which would simply reproduce that axiom as the consequence.  
Finally, if a consequence has terms with derivatives having more than one dependent variable (e.g., $\frac{dx}{dt}$ and $\frac{dy}{dt}$ both appear in the same equation), then we reject the consequence for reasons of numerical solvability, which we discuss in Section~\ref{consequence_data_gen}. We repeat the process until we find an appropriate consequence or run out of choices for variables, derivatives, and constants to add, in which case we repeat with a new sampling. If no consequence is found within $10$ attempts, we discard the axiom system. Note that while computing Gr\"{o}bner basis is known to be doubly exponential in the number of variables in the worst case, in practice Gr\"{o}bner base implementations can be efficient for generic cases \cite{BARDET201549}. The algorithm we run is described in detail in Appendix~\ref{app:consequences}.

\subsection{Data Generation for Consequences}\label{consequence_data_gen}

Now that we can derive a consequence $q(\mathbf{v},\boldsymbol{\delta},\boldsymbol{\lambda})$ from an axiom system $\{f_1,...,f_n\}$,  we next explain the generation of data associated with the given consequence.
Throughout, our sampling will be performed in an independent and identically distributed (i.i.d.) manner, unless noted otherwise.

Given consequence $q(\mathbf{v},\boldsymbol{\delta},\boldsymbol{\lambda})$ obtained from $\{f_1,...,f_n\}$, we restrict the generated consequences to have at most one named physical constant. We are thus able to follow the practice used in AI Hilbert \cite{ai-hilbert24} and assume a scaling of physical units for the physical constants to assume a small value for numerical stability and set the value of this constant to $1$. We do not scale mathematical constants such as $\pi$ or $e$ and instead use their true values.

Next, we sample all measured variables except derivatives and their 
dependent variables (e.g., if the derivative is $\frac{dx}{dt}$ then the dependent variable is $x$)
from a uniform distribution over a random region $[n,m]$, where $n,m$ are random integers in $[1,10]$ with $n<m$. This follows the synthetic data generation approach used by AI Feynman~\cite{doi:10.1126/sciadv.aay2631} and AI Hilbert~\cite{ai-hilbert24}, where the authors generated uniform and normal data with means in $[1,10]$ and $[0,5]$, respectively. \textcolor{black}{If any of $\{\sin\_\theta, \cos\_\theta, e\_\theta\}$ appears among the measured variables, we draw $\theta$ 
from the same base distribution and deterministically compute $\sin\_\theta := \sin(\theta)$, $\cos\_\theta := \cos(\theta)$, $e\_\theta := e^{\theta}$.} 

For consequences containing derivatives, we use an ODE-solver to generate consistent data. When a derivative appears in the consequence polynomial and its dependent variable is among the measured variables, we first solve the consequence polynomial symbolically for the highest order derivative to obtain an expression of the form $\frac{dx}{dt} = f(x, \text{other variables})$ or $\frac{d^2x}{dt^2} = f(x, \frac{dx}{dt}, \text{other variables})$. We then numerically integrate this ODE using SciPy's \texttt{solve\_ivp} Runge-Kutta solver (RK45 method) \cite{2020SciPy-NMeth} with random initial conditions sampled, as earlier,
from the region $[n,m]$ for $n<m$, and $n,m$ random integers in $[1,10]$ consistent with the variable distributions used in \cite{udrescu2020aifeynmanphysicsinspiredmethod}. As mentioned in Section~\ref{Projection exposition}, we limit the number of dependent variables among derivative terms in the consequence to at most one, and thus the solver always deals with an ODE that is numerically solvable. 
If no real solutions exists, we reject the theory as having no consistent solution within the tested region.
For each data point, we sample values for all other measured variables, integrate the ODE over a unit time interval $t \in [0,1]$, and extract a random point from the trajectory (avoiding the initial transient). This produces consistent pairs of (base variable, derivative) values that exactly satisfy the differential relationship encoded in the consequence.

For consequences without derivatives, we generate data for all but one variable and solve for the remaining variable using NumPy's \texttt{roots()} function, taking the first real root of the polynomial after substituting the sampled values. This approach aligns with prior works that sample from distributions for all but one variable and solve for the final variable \cite{doi:10.1126/sciadv.aay2631, ai-descartes23, ai-hilbert24}. Note that when $\text{deg}_q(v_h) \geq 5$, NumPy approximates the roots numerically.

We iterate this process until we obtain 1000 valid data points that satisfy all numerical constraints (real-valued solutions, finite values, within reasonable physical bounds). Finally, we add Gaussian noise to each non-constant column, sampled 
from $N(0, \epsilon \cdot \overline{v})$ where $\overline{v}$ is the column mean, with noise levels $\epsilon \in \{10^{-3}, 10^{-2}, 5 \times 10^{-2}, 10^{-1}\}$.

Our final dataset consists of $216$ dimensionally consistent axiom systems, each of which includes both noiseless and noisy data. 
Hence, the first 
SR
task is: given the dataset and optionally the theory, find the correct consequence that matches the data and ideally is explained by the theory.

\section{Generating Theories with Mismatching Data}\label{Alternate Theories}

\begin{figure} [t]
    \centering
    \includegraphics[scale=0.2]{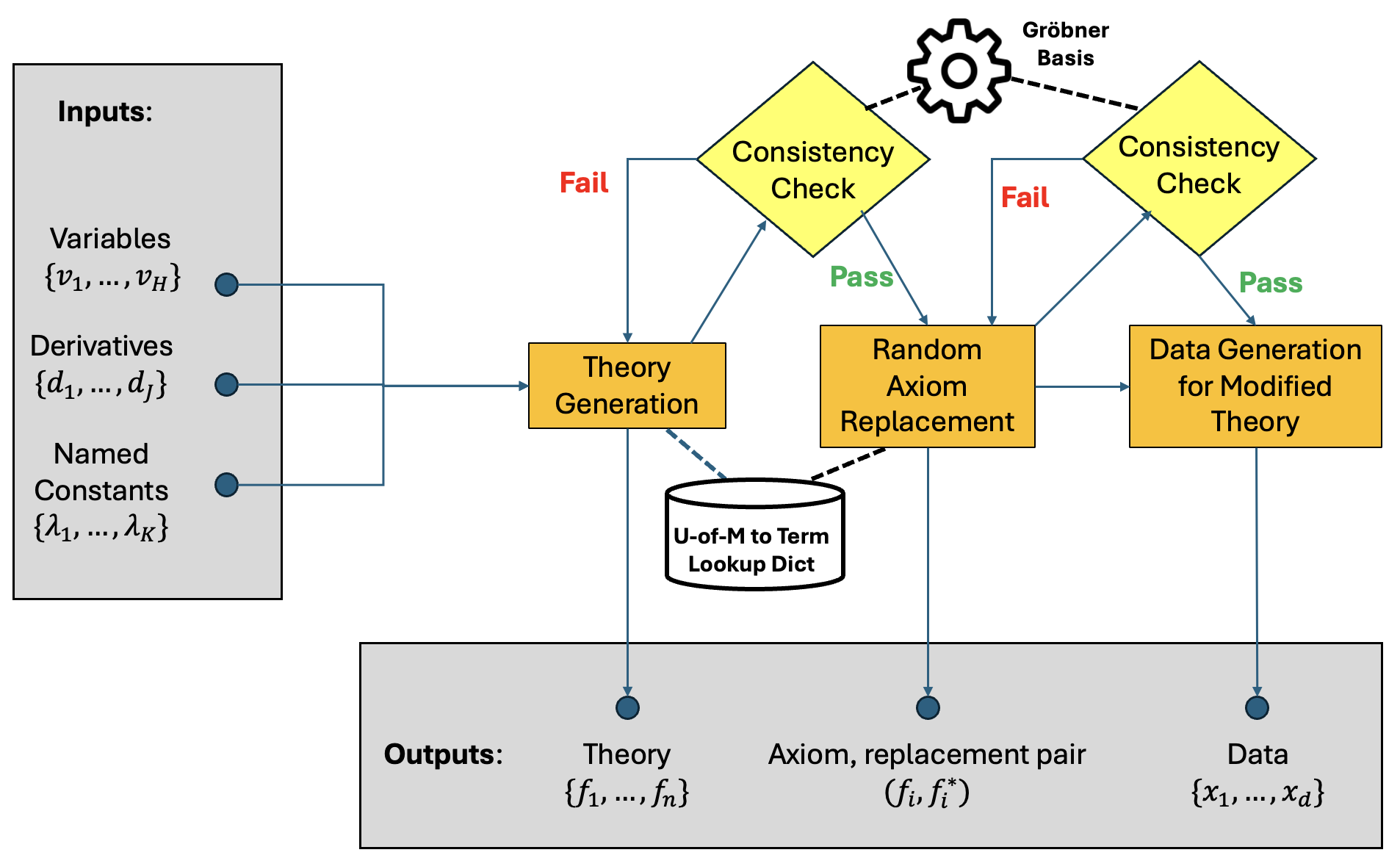}
    \caption{The Pipeline for Generating Theories Together With Mismatching Data.}
    \label{fig:biased-axiom-pipeline}
\end{figure}

In this section, we discuss the case where some portion of a theory is not a good match to select experimental data. 
The discovery and regression challenges are twofold: 
(i) given a potentially incorrect theory and data for a phenomenon, find a governing equation that fits the data well; and (ii) given data consistent with a correct theory but not well explained by a faulty one, identify the flawed axiom(s) and propose one or more replacement axioms that better align with the data. 
The latter challenge is currently beyond the state of the art for  systems that incorporate background theories into the SR task.
We focus our discussion on the case where there is one offending axiom~--~the case with more than one offending axiom is almost entirely analogous. 
A schematic diagram of this pipeline is given in Figure \ref{fig:biased-axiom-pipeline}.

In order to generate a theory with biased data in this sense, we start by generating a theory.\footnote{Alternatively, we could start with a known existing theory.} Then, we randomly replace one of the axioms with a randomly generated replacement axiom such that the resulting system is still consistent, but for which the solution set associated with the new system is different from the solution set associated with the original system, thus capturing different theories. This new system with a replaced axiom will represent a faulty theory while the original will represent the ground truth  system. This replacement has some constraints. For example, consider the three defining equations from which Kepler's Third Law of planetary motions is derivable:
\begin{eqnarray}
    &&d_1m_1 = d_2m_2, \label{k1}\\
    &&F_gd_1^2 + 2F_gd_1d_2 + F_gd_2^2= Gm_1m_2, \label{k2}\\
    && F_g = m_2d_2w^2. \label{k3}
\end{eqnarray}
These are the Newtonian equations that govern two bodies with respective masses $m_1, m_2 $ each of respective distances  $d_1, d_2 $ to their center of mass. $F_g$ is the gravitational force exerted on each body and $w$ is the so-called frequency of the orbit of the second mass, i.e., $w = 1/p$, where $p$ is the time taken by the mass to complete one revolution around the center of mass. Equation (\ref{k3}) comes from the fact that the centrifugal force  $F_c = m_2d_2w^2$  and the observation that $F_c = F_g$. 
Now, suppose we ask SynPAT to replace a random axiom of this system and it selects the third of the three axioms, i.e., equation (\ref{k3}). Since the original axioms are dimensionally consistent, SynPAT will automatically generate a dimensionally consistent replacement axiom. An additional constraint for the generation of the replacement axiom is that it must also contain the variable $w$, since $w$ does not appear in any axiom other than the third axiom. A selection of replacement axioms generated by SynPAT is given by:
\begin{eqnarray*}
    &&F_gd_1 + d_1^2m_1w^2 = F_gd_2  + d_2^2m_1w^2, \\
    &&F_gm_1 = d_1m_1^2w^2 + d_1m_2^2w^2,\\
    && F_g + 2d_1m_1w^2 = d_2m_2w^2. 
\end{eqnarray*}

We then generate data (noisy or noiseless, per specification) for the ground truth theory. Once we have generated an axiom system $\{f_1,\ldots,f_n\}$ as described in Section~\ref{sec:Theory-Generation}, we generate data for the original system by applying the elimination theorem described in Section~\ref{Projection exposition}. Specifically, we compute a Gr\"{o}bner basis $\mathcal{G} = \{g_1,\ldots,g_m\}$ using a lexicographic ordering of the indeterminates. The basis is therefore structured in a way that progressively eliminates indeterminates. This is necessary because, unlike in Section \ref{consequence_data_gen}, there is no phenomenon or consequence and we are generating data that corresponds to the entire axiom system, not just one consequence. We proceed by iteratively following a process similar to that in Section \ref{consequence_data_gen}. For each equation $g_i$ in $\mathcal{G}$, we generate data for all but one indeterminate, setting the at most one constant to the value $1$, and sampling derivatives and variables identically as in Section \ref{consequence_data_gen}, skipping indeterminates for which data have already been sampled. 
As earlier, we use the function \code{roots()} to find the roots of the last variable and take the first real root. The final dataset contains a list of values for all variables in the axiom system. 
Additionally, we add Gaussian noise to each of the variables $v_r$ and derivatives $d_s$ respectively sampled from $N(0,\epsilon \cdot \overline{v}_r)$ 
and
$N(0,\epsilon \cdot \overline{d}_s)$,
where $\overline{v}_r$ is the mean of the data generated for $v_r$ and similarly for $\overline{d}_s$. 
We also have noise level $\epsilon$ values ranging from $10^{-4}$ to $10^{-1}$.

For each of the $216$ systems $\mathcal{F} = \{f_1,...,f_n\}$ generated with corresponding consequences and data in the dataset, we generate both noiseless and noisy data as described above that corresponds to the axiom system as well as five alternate axiom systems $\mathcal{F^*} =\{f_1,...,f_i^*,...,f_n\}$ where a single axiom from the original system has been replaced. 
The dataset, therefore, is well set up to tackle the two previously mentioned regression challenges associated with incorrect theories: discovering a phenomenon from an incorrect theory and data corresponding to the phenomenon, and correcting a faulty theory given data consistent with the correct axiom system. Although current systems are not capable of solving the latter type of problem, 
in the following section we test AI Hilbert on the former type of problem along with benchmarking our dataset for the consequence discovery task described in Section \ref{consequence_data_gen}.

\section{Benchmarking Consequence Discovery}\label{benchmarking}
In this section, we describe the benchmarking of various SR systems on our dataset. We use five systems in our comparison: PySR, AI Feynman, Bayesian Machine Scientist, GPG, and AI Hilbert. 
The first four
are each provided with consequence data on the measured variables, measured derivatives and observed constants $\{\mathbf{v},\boldsymbol{\delta},\boldsymbol{\lambda}\}$, and tasked with predicting $v_h$ in the relation 
\[
v_h \approx f(v_1, ..., v_{h-1}, \boldsymbol{\delta}, \boldsymbol{\lambda}),
\]
where $v_h$ is obtained from the first real root of a univariate polynomial determined by the theory. AI Hilbert, in addition, is given the full set of axioms $\{f_1, ..., f_n\}$ along with the data. We further test AI Hilbert on the alternate axiom systems from Section \ref{Alternate Theories}, where one axiom is replaced with an incorrect axiom insufficient to prove the consequence. 

We sampled $44$ systems across three independently varying structural parameters:
\textbf{Number of variables} $\in \{6,7,8,9\}$, \textbf{Number of derivatives} $\in \{2,3,4\}$, \textbf{Number of equations (axioms)} $\in \{4,5,6\}$. Note that for each configuration, we evaluate model performance at three noise levels: $\epsilon \in \{0, 10^{-3}, 10^{-1}\}$, as described in Section \ref{consequence_data_gen}. The results are presented in 
Tables~\ref{tab:models-comparison_trig}~--~\ref{tab:eq-performance_trig}
in Appendix \ref{Benchmarking Results}, broken down by the number of variables, derivatives, and equations, respectively. A summary of the results is given in Table~\ref{tab:summary-benchmark}.

\begin{table}[h] 
\footnotesize
    \begin{center}
     \caption{Number of Correct Expressions out of 36 Systems for Various Methods at Varying Noise Levels.}
    \renewcommand{\arraystretch}{1.2}
    \begin{tabular}{|c|c|c|c|}
        \hline
        {\textbf{Method}} & \multicolumn{3}{c|}{\textbf{Noise Level} ($\epsilon$)} \\ 
        \cline{2-4}
        & \bm{$0$} & \bm{$10^{-3}$} & \bm{$10^{-1}$} \\ 
        \hline
        PySR & 15 & 9 & 1 \\ 
        \hline
        AI Feynman & 16 & 3 & 1 \\ 
        \hline
        AI Hilbert & 23 & 17 & 8 \\ 
        \hline
        AI Hilbert w/ replaced axioms & 12 & 7 & 3 \\ 
        \hline
        Bayesian Machine Scientist & 10 & 5 & 3 \\ 
        \hline
        GPG & 14 & 5 & 1 \\ 
        \hline
    \end{tabular}
    \vspace*{-10pt}
    \label{tab:summary-benchmark}
\end{center}
\end{table}

\noindent The results demonstrate that AI Hilbert outperforms the other systems across all noise conditions, achieving $64\%$ accuracy on noiseless datasets and maintaining $22\%$ accuracy even at high noise levels ($\epsilon = 10^{-1}$), suggesting that leveraging background theory can be advantageous for machine-assisted discovery of physical theories. The newly benchmarked methods, Bayesian Machine Scientist and GPG, show moderate performance relative to the established baselines, with GPG achieving $39\%$ accuracy at zero noise but degrading more sharply with increasing noise. We also observe that providing AI Hilbert with a flawed background theory notably decreases performance from $64\%$ accuracy on noiseless datasets to $33\%$ accuracy on the same and from $47\%$ to $8\%$ at a noise level $\epsilon = 10^{-1}$, demonstrating that the correctness of background theory has a direct and substantial impact on its performance.
A more detailed breakdown and discussion of performance based on the various parameters can be found in Appendix \ref{Benchmarking Results}. 

SR systems sometimes assume, either implicitly or explicitly, a Gaussian noise distribution. Gaussian noise tends to be a good model of random measurement error but a less good model of systematic error  associated with the measurement device itself (including erroneous assumptions about what the measuring device is in fact measuring). With this realization in mind, we have performed an analysis to understand how sensitive the various SR systems in this study are to the noise \emph{type} as well as the noise quantity. It turns out that the SR systems are not particularly sensitive to the noise type at low noise levels, but are more so at high noise levels. See Appendix \ref{sec:sensitivity-analysis} for a detailed analysis and discussion.

\section{Code and Generated Data Set Availability}\label{CodeAvail}

All code and scripts needed to reproduce the results of this paper, together with generated data sets, and artifacts used to parameterize the SynPAT system (including our analysis of equations from the Feynman database \cite{f-srdb2020}) are freely available under an open source MIT license. The implementation allows for adjustment of the number of variables, constants, derivatives, and equations (as well as their specification). In addition, one can specify variable and derivative value ranges and specify the parameters governing noisy data models. We provide complete documentation of these and more configurable parameters and their effects on theory generation in the repository. A detailed breakdown of the computational resources used in generating the dataset is provided in Appendix \ref{computation_resources}. 

\section{Conclusions and Future Work}

\textcolor{black}{
Our system of decoupling transcendental semantics from algebraic reasoning by introducing dimensionless auxiliary symbols \(\sin\_\theta,\cos\_\theta,e\_\theta\) (with \(\theta\) also dimensionless) that behave like polynomial indeterminates as well as our ability to similarly generate consequences for axiom systems with ordinary differential equations allows us to generate axioms and consequences analogous to the vast majority of the equations in the Feynman dataset. Our framework is adaptable and can be used to incorporate any special functions whose characteristics can be captured by polynomial axioms and data. For example, one could extend the framework to build in a function like $\ln(x)$ and one would only need to adjust the data generation process to account for this function. A limitation is that non-algebraic relationships between indeterminates cannot be axiomatized, so that we cannot express $e^{\ln(x)}=x$, since exponentiation is not an inherent part of the algebraic Gr\"{o}bner basis formalism.}  

Our benchmarking of 
SR
systems, some with the ability to incorporate an existing theory and some without, is not meant to serve as an exhaustive comparison of such systems. Rather, the benchmarking is meant to serve as an examplary usage of the SynPAT system, exhibiting its wide-ranging functionality. 
Finally, it should be noted that there is one use case for which we have generated data that no current system can solve: namely, cases where we start with a flawed theory, replace a random axiom with a randomly generated replacement that is inconsistent with the original system, generate data for this new system, and ask the regression systems to identify the flawed axiom along with a suitable replacement that better explains the data. This is the challenge that we believe systems capable of theory-based reasoning should aim to solve.

\bibliography{references}
\bibliographystyle{icml2026}

\newpage
\appendix
\onecolumn

\section{Theory Generation Algorithms} \label{app:theory-gen-algs}

\subsection{Term Generation} \label{app:term-gen}
The first component of theory generation is the algorithm to generate a random term. Random terms are generated as follows. First, a random number of factors is determined -- a factor being a particular variable, derivative, or named constant of those specified by the user. These probabilities are again gleaned from our compiled list of uniformized Feynman equations and are as follows: $[0.246, 0.413, 0.225, 0.088, 0.025, 0.004]$, meaning that there is a $0.246$ probability that the term will contain a single factor, a probability of $0.413$ that a term will contain exactly two factors, and so on. 
For each possible factor, there is the possibility that the factors will appear raised to an integral exponent greater than 1. The probability of each factor being raised to each possible power is again mediated through parameters that sum to unity. Variables and named constants are far more likely to be raised to powers beyond $1$ than derivatives. Following the assignment of powers to the various factors, an unnamed multiplicative constant is assigned. Of the 96 Feynman equations that we are able to work with, none have \emph{non-integer} unnamed constants. There is one instance of the constant 20 appearing, another where the constant 32 appears, and a single instance where the constant 6 appears. If we ignore these as ``sporadic'' constants, the relative frequency of the implicit constant 1 together with the constants 2, 3, and 4 are as follows: $[0.82, 0.07, 0.06, 0.05]$, which are the probabilities we use to assign these constants. We note in passing that there is really no computational cost associated with including the constants 6, 20 and 32 as possibilities, if users of our code so desire -- they can simply be added to the array named $\mathtt{PROB\_OF\_SMALL\_INTEGER\_CONSTANTS}$ in the \textbf{Equation} class.
All small integer constants are assumed to be dimensionless. The selection of factors and powers of the factors, together with the unnamed constant multiplier, completely determine the term. One important nuance in term generation is that terms may not have more than one distinct derivative factor, meaning that, e.g., there are no terms of the form $\frac{dx}{dt}\frac{d^2x}{dt^2}$, or of the form $\frac{dx}{dt}\frac{dy}{dt}$, in each case, possibly with some additional factors. Such mixed ordinary derivative terms seem never to appear in extant physical systems. (Refer to ODE Bench mentioned in \cite{d'ascoli2024odeformer} and the ODE Strogatz repository \cite{stro-repo2020}.) A term may also have a positive or negative sign within an equation, but these are generated for all terms at once within the equation generation routine.

The direct inputs to the term generation algorithm are the sets $\mathcal{V}$ of variables, $\mathcal{D}$ of derivatives, and $\mathcal{C}$ of constants, along with $\mathtt{max\_power}$, the maximum power to which to raise any variable, derivative, or constant. The other variables listed as inputs to this algorithm are class variables for the \textbf{Equation} class. We adopt the convention of using variable names with all capital letters to designate class variables that hold essential parameters of the model. Pseudocode for term generation is shown in Algorithm \ref{alg:term-gen}.

\begin{algorithm} 
\caption{Generate Random Term}
\begin{algorithmic}[1]
\REQUIRE
Sets $\mathcal{V}$ of variables, $\mathcal{D}$ of derivatives, and $\mathcal{C}$ of constants; $\mathtt{max\_power}$ the maximum power to which to raise any variable, derivative or constant; $\mathtt{PROB\_FACTORS\_PER\_TERM}[]$ an array giving the probability of each number of different factors per term for number of factors = 1, 2, 3 and 4; $\mathtt{SAME\_FACTOR\_VARIABLE\_BIAS}$ a real number giving likelihood that a variable or named constant gets raised to a power $> 1$; $\mathtt{SAME\_FACTOR\_DERIVATIVE\_BIAS}$ a real number giving likelihood that a derivative gets raised to a power $> 1$; and $\mathtt{PROB\_OF\_SMALL\_INTEGER\_CONSTANTS}[]$ an array giving the probability of each small integer multiplicative constant for the values 1, 2, 3 and 4.
\ENSURE
A random term using $(\mathcal{V}, \mathcal{D}, \mathcal{C})$.
\STATE From $\mathtt{PROB\_FACTORS\_PER\_TERM}[]$:
\\
\ \ \ \ \ \
 $\mathtt{num\_factors} \gets$ randint$(1,..,4)$ 
\STATE $\mathtt{factors} \gets \mathtt{num\_factors}$ random elements from $\mathcal{V} \cup \mathcal{D} \cup \mathcal{C}$
\STATE $\mathtt{term} \gets 1$
\FOR{$\mathtt{factor}$ in $\mathtt{factors}$}
    \IF{$\mathtt{factor}$ isinstance(Variable) or  isinstance(Constant)}
        \STATE From $\mathtt{SAME\_FACTOR\_VARIABLE\_BIAS}, \mathtt{max\_power}$:
        \\
        \ \ \ \ \ \ \ \  \ \ \ \ \ \ \ \ \ \ pow $\gets$ random exponent
    \ELSE
        \STATE From $\mathtt{SAME\_FACTOR\_DERIVATIVE\_BIAS}, \mathtt{max\_power}$:
        \\
        \ \ \ \ \ \ \ \  \ \ \ \ \ \ \ \ \ \ pow $\gets$ random exponent
    \ENDIF
    \\
    \ \ \ \ \ \ $\mathtt{term} \gets \mathtt{term}$ * $\mathtt{factor}$ ** $\mathtt{pow}$
\ENDFOR
\STATE From $\mathtt{PROB\_OF\_SMALL\_INTEGER\_CONSTANTS}[]$:
\\
\ \ \ \ \ \ $\mathtt{c} \gets$ randint$(1,..,4)$
\STATE $\mathtt{term} \gets \mathtt{c}*\mathtt{term}$
\STATE \textbf{Return} $\mathtt{term}$
\end{algorithmic}
\label{alg:term-gen}
\end{algorithm}

\subsection{Equation Generation} \label{app:eqn-gen}

The second component of theory generation is the algorithm to generate either random equations, irrespective of the dimension (unit of measure) of the terms, or random dimensionally consistent equations. We first describe the generation of random, not necessarily dimensionally consistent equations.

\subsubsection{Generating Random Not-Necessarily Dimensionally Consistent Equations} \label{app:eqn-gen-not-dim-consistent}

The first step in the generation of such an equation is generating a random number of terms. From the 100 equations in the Feynman lecture series database \cite{f-srdb2020}, we compiled a list of the 70 equations that match our model (mostly those without special functions
), and uniformized them so that they only deal with one spatial dimension
, and from these equations we found the distribution of the number of terms per axiom to be $[0, 0.6, 0.29, 0.09, 0.03]$, meaning there is a $0$ probability that an equation has just one term, $0.6$ probability that it has two terms, and so on. 

Once the random number of terms are generated, three things are checked: (i) that no two terms are equal modulo constants; (ii) that the terms do not all contain a common factor; and (iii) that there are not exactly two terms, each of which is a monomial. The latter condition means that one of the variables, derivatives, or named constants is redundant. In either of these three cases, the existing set of terms is discarded and a new set is regenerated
. Once a good set of terms is obtained, random signs are assigned to the terms. There is one caveat, however. It is relatively rare in existing theories that one comes across an equation of the form $t_1 + \cdot\cdot\cdot + t_k = 0$. The probability of such an occurrence is set to 0.05 -- in which case all terms are given a positive sign. In all other cases, terms are successively assigned random signs until the last term. If all prior terms have been given the same sign, this last term is automatically given the opposite sign. In the event that all prior terms have \emph{not} been given the same sign, this last term is also given a random sign. With the assignment of random signs, the definition of the equation is complete. In the case that a non-dimensionally-consistent set of equations has been requested, the given number of equations are all generated in the same manner. After all equations are generated, a test is run to guarantee that no two generated equations are equal (an extreme rarity). 

\subsubsection{Generating Random Dimensionally Consistent Equations} \label{app:eqn-gen-dim-consistent}  

The first step in the process of generating a dimensionally consistent equation is to determine an acceptable $\mathtt{max\_power}$, or maximum power to which the generator will be willing to raise monomials (the candidate variables, derivatives, and named constants).  The key parameters in the overall process are $\mathtt{max\_power}$ together with $\mathtt{num\_monomials}$ (the number of distinct candidate variables, derivatives, and named constants) and $\mathtt{max\_factors}$ (the maximum number of distinct candidate variables, derivatives and named constants appearing in any term). We will see that the runtime of the generation process is 
\begin{equation} \label{runtime} O\Bigg((\mathtt{max\_power}+1)^{\mathtt{max\_factors}} * \binom{\mathtt{num\_monomials}} {\mathtt{max\_factors}}\Bigg),
\end{equation}
which is manageable for the range of values we consider. In typical theories, 
$\mathtt{num\_monomials}$ ranges from $8$ to $20$, while, recalling from the probabilities we arrived at for the number of factors from the Feynman equations, we settled upon $\mathtt{max\_factors} = 4$ even though there were also very small probabilities for $\mathtt{max\_factors} \in \{5,6\}$. We have generally set $\mathtt{max\_power} = 3$, though there are certainly equations in the Feynman set where $\mathtt{max\_power}$ is higher. Thus, for our generated theories, the quantity inside the big $O()$ of expression (\ref{runtime}), comes to $4^4 * {20 \choose 4 } = 256*4845 = 1,240,320$. To keep our simulations moving fast, we sometimes set $\mathtt{max\_factors} = 3$ so long as $\mathtt{num\_monomials}$ does not exceed some critical value, e.g., for values of $\mathtt{num\_monomials} < 15$. At the outer limits of what we see in the Feynman dataset, if we were to set $\mathtt{max\_factors} = 6, \mathtt{max\_power} = 6$, we would obtain
\begin{equation*}
    (\mathtt{max\_power}+1)^{\mathtt{max\_factors}} *{\mathtt{num\_monomials} \choose{\mathtt{max\_factors}}} = 7^6 * {20 \choose 6}
    = 4,560,075,240, 
\end{equation*}
which is not practical, especially since the constant implicit in (\ref{runtime}) is substantial.

Once the value of \texttt{max\_power} is set, two dictionaries are created. The first is a unit-of-measure-to-primitive-term lookup dictionary. By ``primitive term'' we mean a term with an implicit constant of~1 and positive sign. The idea is that a random first term will be created, and then additional terms will be added, all of which must have the same unit of measure. The unit of measure for the initial term will be determined, and the remaining terms will be selected from this unit-of-measure-to-primitive-term lookup dictionary. There are precisely

\vspace*{-\baselineskip}
\begin{equation*}
{\mathtt{num\_monomials} \choose 1} + 
{\mathtt{num\_monomials} \choose 2} + \cdot\cdot\cdot +
{\mathtt{num\_monomials} \choose \mathtt{max\_factors}}
\end{equation*} 

ways to choose up to $\mathtt{max\_factors}$ factors. Each of these factors must be assigned a power from $1$ to $\mathtt{max\_power}$. The result is a primitive term. To create the unit-of-measure-to-primitive-term lookup dictionary, we run sequentially through all such primitive terms, each time computing the unit of measure. For each pair of the form $(\mathtt{unit~of~measure},~ \mathtt{primitive~term})$, the in-process-of-being-created dictionary is consulted. If there is not yet an entry for the given unit of measure, a new entry is created with said unit of measure as key and a value that is a single element list consisting of the given primitive term. If there is already an entry for the given unit of measure, the primitive term is appended to the list of primitive terms for that unit of measure. After cycling through all primitive terms, the dictionary is created. 

The second dictionary that is created is what we call the vdc (for variable-derivartive-constant) signature distribution dictionary. The idea behind this dictionary is that for a given set of variables, derivatives and constants, not all terms comprising these variables, derivatives and constants are equally likely to be encountered, and analogously, the set of possible terms that are consistent with a given unit of measure are not equiprobable. For example, for a given set of, say four variables, derivatives and constants comprising a particular term, it is relatively unlikely that each factor will be raised to the $\mathtt{max\_factor}$. Moreover, derivatives are more unlikely than variables to be raised to high powers. The so-called ``vdc signature'' is the key to this second dictionary and something that can be computed based on the powers that each variable, derivative and constant is raised to in a given term. The idea of a vdc signature is that the probability of occurrence of two terms with the same vdc signature should be roughly the same. The vdc signature is a string comprising three components separated by periods (so a little bit like the Dewey Decimal System for cataloging books in libraries). The first segment of the signature identifies the distribution of powers among any constituent variables, the second segment identifies the distribution of powers among constituent derivatives, and the last segment identifies the distribution of powers among constituent constants. For example, if $x,y,z$ are variables and $c$ is a constant, the term $c*x*y**2*z~ (=cxy^2z)$ has the signature ``112..1'', where the leading segment `112' refers to the fact that there are three distinct variables in the term and they are raised to the respective powers $1,1$ and $2$. The fact that there are no numbers between the two decimal points means there are no derivatives appearing in the term, and the final $1$ signifies the fact that there is one constant appearing and it is raised to the first power. The numbers 112 appearing in the first segment appear in this order because the convention is that the numbers within a segment are sorted from smallest to largest. For another example, consider the case of a term featuring the constant $G$, and the derivatives $dxdt~(=\frac{dx}{dt}), dydt~(=\frac{dy}{dt})$, of the form $G{**}2*dxdt{**}2dydt~(=G^2(\frac{dx}{dt})^2\frac{dy}{dt})$. The vdc signature for this term is ``.12.2'', where the lack of an initial segment means there are no variables in the term, the `12' middle segment means there are two derivative factors appearing, one raised to the 1st power, the other raised to the 2nd power, and the final '2' indicates that there is a single constant raised to the 2nd power. With an understanding of how the vdc signature is created, the vdc signature distribution dictionary is created by generating 10,000 random terms (in accordance with how we described term generation to be run, following the various probabilities as gleaned from the AI Feynman database), computing the vdc signature of each term so generated, these signatures becoming the dictionary keys, and maintaining a frequency count for each such signature~--~with the final counts becoming the dictionary values.

Once the unit-of-measure-to-primitive-term dictionary and the vdc signature distribution dictionaries have been assembled, we are ready for the formal parts of dimensionally consistent equation (and equation system) generation to begin. First, a random number of terms, which we henceforth denote by $\mathtt{num\_terms}$, is determined, as well as a random first term -- both obtained precisely as was done when we were generating a random equation without regard to dimensional consistency. The unit of measure associated with the first term is then obtained. For example, if the term is  $\frac{d^2x}{dt^2}Fm$ where $\frac{d^2x}{dt^2}$ is an acceleration, $F$ is a force and  $m$ is a mass, then the respective reduced units of measure\footnote{Meaning, not including any derived units of measure, such as N (Newtons = kg*m/s$^2$).} are, respectively, $m/s^2, kg*m/s^2, kg$ so that the aggregate unit of measure for the term is the product of all of these, or $kg^2 * m^2/s^4$. All further terms for this equation must have this same aggregate unit of measure. Once the unit of measure is obtained, the viable primitive terms for the unit of measure are fetched from the unit-of-measure-to-primitive-term dictionary. If there are fewer than $\mathtt{num\_terms}$ matching terms for this unit of measure, a new random $\mathtt{num\_terms}$ and first term are generated. Finally, once a satisfactory $\mathtt{num\_terms}$ and first term are generated, the remaining $(\mathtt{num\_terms} - 1)$ primitive terms are drawn at random from the pool of possible terms but observing the relative frequency distribution for the associated vdc signatures as gleaned from the vdc signature distribution dictionary. In other words if $(\mathtt{num\_terms}-1) = 2$ and there are four candidate terms with the same unit of measure $t_1,t_2, t_3,t_4$, corresponding vdc signatures $v_1, v_2, v_2, v_3$, and associated frequency counts for three signatures $(v_1, v_2, v_3)$ of, respectively $(25, 50, 100)$, then the two terms would be picked from the four candidates, each with associated probability $\frac{1}{9}, \frac{2}{9}, \frac{2}{9}, \frac{4}{9}$, where, e.g., the probability of picking $t_1$ is $\frac{1}{9} = \frac{25}{25 + 50 + 50 + 100}$. The assignment of non-unital constants is performed in the same way as was done when we were generating a random equation without regard to dimensional consistency. The same two sanity checks are then performed as earlier: that there are not exactly two terms, each containing a single indeterminant, and that all the terms do not have a common factor. If either of these sanity checks fails, the process is repeated, starting with the generation of a random number of terms, $\mathtt{num\_terms}$, and a random first term. 

Since generating random equations, irrespective of the dimension of the terms, is similar to generating random dimensionally consistent equations, except that most of the steps are omitted, we present just the pseudocode for the dimensionally consistent case -- see Algorithm \ref{alg:eqn-gen}. This algorithm is invoked repeatedly when generating random dimensionally consistent \textbf{EquationSystem}s. Two of the parameters it is invoked with are the two dictionaries mentioned in the main text: (i) the unit-of-measure-to-term lookup dictionary ($\mathtt{u\_of\_mToTermLookupDict}$), and (ii) the vdc (variable-derivative-constant) signature distribution dictionary ($\mathtt{vdcSigProbDict}$). As discussed in the main text, the creation of the unit-of-measure-to-term lookup dictionary is the main bottleneck of the theory generation system and thus when many \textbf{EquationSystem}s are created for the same sets of variables, derivatives, and constants, this lookup dictionary is stored and reused. It is stored as a class variable in the \textbf{EquationSystem} class and also generated as a class method there. On the very first call to Algorithm \ref{alg:eqn-gen} the dictionary is created and forever after it is fetched from the \textbf{EquationSystem} class variable in the invocation of the algorithm and thereby reused.   Although less time consuming to create, the same create once and reuse process is applied for the vdc signature distribution dictionary, which is similarly stored as a class variable in the \textbf{EquationSystem} class.

\begin{algorithm} [t] 
\caption{Generate Random Dimensionally Consistent Equation}
\begin{algorithmic}[1]
\REQUIRE 
Sets $\mathcal{V}$ of variables, $\mathcal{D}$ of derivatives, and $\mathcal{C}$ of constants; $\mathtt{max\_power}$ the maximum power to which to raise any variable, derivative or constant; $\mathtt{u\_of\_mToTermLookupDict}$ unit-of-measure-to-term lookup dictionary; $\mathtt{vdcSigProbDict}$ vdc signature probability lookup dictionary; $\mathtt{PROB\_OF\_NUM\_TERMS}[]$ an array giving the probability of an equation having each of the respective number of terms equal to 2, 3, 4 and 5; and $\mathtt{PROB\_OF\_SMALL\_INTEGER\_CONSTANTS}[]$ an array giving the probability of each small integer multiplicative constant for the values 1, 2, 3 and 4.
\ENSURE 
A random dimensionally consistent equation using $(\mathcal{V}, \mathcal{D}, \mathcal{C})$.

\IF{$\mathtt{u\_of\_mToTermLookupDict}$ is $\mathtt{None}$}
        \STATE $\mathtt{u\_of\_mToTermLookupDict} \gets$  
        GenerateUofMToPrimitiveTermLookupDic(
        $\mathcal{V}, \mathcal{D}, \mathcal{C}, \mathtt{max\_power}$) 
\ENDIF
\IF{$\mathtt{vdcSigProbDict}$ is $\mathtt{None}$}
        \STATE $\mathtt{vdcSigProbDict} \gets$  
        GenerateVDCSigProbDict($\mathcal{V}, \mathcal{D}, \mathcal{C}, \mathtt{max\_power}$)
\ENDIF
\WHILE{True}
    \STATE From $\mathtt{PROB\_OF\_NUM\_TERMS}[]$:
    \\
    \ \ \ \ \ \ \ \ \ \ \ \ 
    $\mathtt{num\_terms} \gets$ randint$(1,..,4)$
    \STATE $\mathtt{firstTerm} \gets$ GenerateRandomTerm($\mathcal{V}, \mathcal{D}, \mathcal{C}, \mathtt{max\_power}$)
    \STATE $\mathtt{u\_of\_m\_to\_match} \gets$ GetUofMForTerm($\mathtt{firstTerm}$)
    \STATE $\mathtt{candidateTerms} \gets \mathtt{u\_of\_mToTermLookupDict}$.get(
    $\mathtt{u\_of\_m\_to\_match}$)
    \IF{$|\mathtt{candidateTerms}| < \mathtt{num\_term}$ \textbf{or} 
    HasCommonFactors($\mathtt{candidateTerms}$) }
        \STATE Continue to next iteration
    \ENDIF
    \STATE $\mathtt{candidateTerms}$.remove(ConstantStrip($\mathtt{firstTerm}$))
    \STATE $\mathtt{vdc\_sig\_frequencies} = []$
    \FOR{$\mathtt{term}$ in $\mathtt{candidateTerms}$}
    \STATE $\mathtt{vdc\_sig} \gets$ GetVDCSigForTerm$(\mathtt{term})$
    \STATE $\mathtt{vdc\_sig\_freq} \gets \mathtt{vdcSigProbDict}$.get($\mathtt{vdc\_sig}$)
    \STATE $\mathtt{vdc\_sig\_frequencies}$.append($\mathtt{vdc\_sig\_freq}$)
    \ENDFOR
    \STATE $\mathtt{additionalTerms} \gets \mathtt{num\_terms} - 1$ random terms from 
    $\mathtt{candidateTerms}$ using $\mathtt{vdc\_sig\_frequencies}$
    \STATE $\mathtt{terms} \gets \{\mathtt{firstTerm}\}$
    \IF{HasCommonFactors($\mathtt{terms}$) \textbf{or} 
    FailsSanityCheck($\mathtt{terms}$) }
        \STATE Continue to next iteration
    \ENDIF
    \FOR{$\mathtt{term}$ in $\mathtt{candidateTerms}$}
        \STATE From $\mathtt{PROB\_OF\_SMALL\_INTEGER\_CONSTANTS}[]$: 
        \\
        \ \ \ \ \ \ \ \ \ \ \ \ $\mathtt{c} \gets$ randint$(1,..,4)$
        \STATE $\mathtt{term} \gets \mathtt{c}*\mathtt{term}$
        \STATE $\mathtt{terms} \gets \mathtt{terms} \cup \{\mathtt{term}\}$
    \ENDFOR
    \STATE DivideByCommonUnnamedConstantFactors($\mathtt{terms}$)
    \STATE AssignRamdomSignsToTerms($\mathtt{terms}$)
\ENDWHILE
\STATE \textbf{Return} Equation($\mathtt{terms}$)
\end{algorithmic}
\label{alg:eqn-gen}
\end{algorithm}

\subsection{Equation System Generation} \label{app:eqn-sys-gen}
The process for generating a random dimensionally consistent set of equations is much the same as that for generating a random not-necessarily dimensionally consistent set of equations, with the exception of course that the call to get a next equation is to a version that generates a dimensionally consistent equation. There are a couple of additional nuances in the dimensionally consistent case, so we focus just on this case. Terms having the same unit of measure have a much higher likelihood of having factors in common than otherwise. Hence, when generating dimensionally consistent sets of equations, special care must be taken to try to use all the variables, derivatives, and named constants. Thus, after each equation is generated, the system makes sure to generate a first term that contains an as yet unused variable, derivative, or named constant, if there are still any that remain unused. An additional sanity check is also performed for this kind of system of equations. There is a check that no two equations have the same unit of measure, unless the two equations use disjoint sets of variables or derivatives. Equations having the same unit of measure and overlapping variables are often inconsistent or force one or more of the variables/derivatives to take the value $0$, which is typically undesirable. 

Pseudocode for this algorithm is provided in Algorithm \ref{alg:eqn-sys-gen}. While the pseudocode shows the two dictionaries, $\mathtt{u\_of\_mToTermLookupDict}$ and $\mathtt{vdcSigProbDict}$ as input parameters, we note that this is very slightly misleading since these dictionaries are class variables in the \textbf{EquationSystem} class and the method for generating random dimensionally consistent \textbf{EquationSystems} is a class method of this same class, so the two dictionaries are directly accessible to the method. Each time the method is called for a particular set of variables, derivatives and constants $(\mathcal{V}, \mathcal{D}, \mathcal{C})$, the set  $(\mathcal{V}, \mathcal{D}, \mathcal{C})$ is stored, together with the generated dictionaries, $\mathtt{u\_of\_mToTermLookupDict}$ and $\mathtt{vdcSigProbDict}$. If the method is called again with the same set $(\mathcal{V}, \mathcal{D}, \mathcal{C})$, the previously generated dictionaries are used and not regenerated. Since we often generate hundreds of \textbf{EquationSystems} for the same sets of variables, derivatives, and constants, this approach saves an enormous amount of runtime compared to regeerating the dictionaries for each \textbf{EquationSystem}. The algorithm uses a very slightly different version of Algorithm \ref{alg:eqn-gen}, namely one where a specified variable, derivative, or constant \emph{must} be included in the generated \textbf{Equation} (see line 16 of the pseudocode). The only difference between this algorithmic variant and that shown in Algorithm \ref{alg:eqn-gen} is that the initial generation of a random term is repeated until the supplied variable, derivative or constant is included. \textcolor{black}{Additionally, when both $\sin\theta$ and $\cos\theta$ are present in the sampled variable set, 
we additionally include the unitless axiom $\sin^2\theta + \cos^2\theta = 1$ to enforce the 
standard trigonometric identity.}

\begin{algorithm} [H]
\caption{Generate Random Dimensionally Consistent Equation System}
\begin{algorithmic}[1]
\REQUIRE 
Sets $\mathcal{V}$ of variables, $\mathcal{D}$ of derivatives, and $\mathcal{C}$ of constants; $\mathtt{numEqns}$ the number of equations to generate; $\mathtt{max\_vars\_derivatives\_and\_constants\_per\_eqn}$ the maximum number of distinct variables, derivatives and constants that may appear in any single equation; $\mathtt{u\_of\_mToTermLookupDict}$ unit-of-measure-to-term lookup dictionary; and $\mathtt{vdcSigProbDict}$ vdc signature probability lookup dictionary.
\ENSURE 
A random dimensionally consistent list of $\mathtt{numEqns}$ equations using as many of $(\mathcal{V}, \mathcal{D}, \mathcal{C})$ as possible.
\STATE $\mathtt{eqns} = []$
\STATE $\mathtt{max\_power} \gets$ DetermineMaxPower$(\mathcal{V}, \mathcal{D}, \mathcal{C})$
\IF{$\mathtt{u\_of\_mToTermLookupDict}$ is $\mathtt{None}$}
        \STATE $\mathtt{u\_of\_mToTermLookupDict} \gets$  
        GenerateUofMToPrimitiveTermLookupDic(
        $\mathcal{V}, \mathcal{D}, \mathcal{C}, \mathtt{max\_power}$) 
\ENDIF
\IF{$\mathtt{vdcSigProbDict}$ is $\mathtt{None}$}
        \STATE $\mathtt{vdcSigProbDict} \gets$  
        GenerateVDCSigProbDict($\mathcal{V}, \mathcal{D}, \mathcal{C}, \mathtt{max\_power}$)
\ENDIF
\STATE $\mathtt{varsDerivsAndConstantsToBeUsed} \gets \mathcal{V} \cup \mathcal{D} \cup \mathcal{C}$ 
\STATE $\mathtt{eqn} \gets$ GenRandomDimConsistentEqn(
$\mathcal{V},\mathcal{D}, \mathcal{C},\mathtt{max\_power}, \mathtt{u\_of\_mToTermLookupDict}$,
$\mathtt{vdcSigProbDict})$
\STATE $\mathtt{eqns}$.append($\mathtt{eqn}$)
\STATE $\mathtt{vdcsUsed} \gets$ GetVarsDerivsAndConstantsUsed($\mathtt{eqn}$) 
\STATE $\mathtt{varsDerivsAndConstantsToBeUsed} \gets$
$\mathtt{varsDerivsAndConstantsToBeUsed} \setminus \mathtt{vdcsUsed}$
\WHILE{$\mathtt{varsDerivsAndConstantsToBeUsed} \neq \varnothing$}
    \STATE $\mathtt{next\_vdc} \gets \mathtt{varsDerivsAndConstantsToBeUsed}$.pop()
    \STATE $\mathtt{eqn} \gets$ 
    GenRandomDimConsistentEqnWithDVC($\mathtt{next\_vdc}$)
    \IF{$\mathtt{eqn}$ is not $\mathtt{None}$}
    \STATE $\mathtt{eqns}$.append($\mathtt{eqn}$)
    \STATE $\mathtt{vdcsUsed} \gets$ GetVarsDerivsAndConstantsUsed($\mathtt{eqn}$) 
    \STATE $\mathtt{varsDerivsAndConstantsToBeUsed} \gets$
    $\mathtt{varsDerivsAndConstantsToBeUsed} \setminus \mathtt{vdcsUsed}$
    \ENDIF
\ENDWHILE
\STATE \textbf{Return} $\mathtt{eqns}$
\end{algorithmic}
\label{alg:eqn-sys-gen}
\end{algorithm}

\section{Consequence Generation and Consistency Check Algorithm} \label{app:consequences}

\begin{algorithm}
\caption{Generating Consequences of Axioms and Checking for Consistency}
\begin{algorithmic}[1]
\REQUIRE 
Axiom system $\mathcal{A}$; set of unknowns $\mathcal{V}$; max attempts $N_{\text{max}}$.
\ENSURE 
Derived polynomial from the Gröbner basis elimination or $\emptyset$ if unsuccessful.

\STATE Initialize $\mathcal{M} \gets \emptyset$ (measured and observed indeterminates)
\STATE Initialize $\mathcal{N} \gets \emptyset$ (remaining indeterminates)
\STATE Set attempt counter $n \gets 0$

\WHILE{$n < N_{\text{max}}$}
    \STATE Shuffle $\mathcal{V}$ randomly
    \FOR{$j = 1$ to $|\mathcal{V}|$}
        \STATE $\mathcal{M} \gets \mathcal{V}[1:j]$
        \STATE $\mathcal{N} \gets \mathcal{V} \setminus \mathcal{M}$

        \IF{$\mathcal{M}$ overlaps axiom subsets in $\mathcal{A}$}
            \STATE \textbf{continue}
        \ENDIF

        \STATE Construct polynomial ring $\mathbb{R}[\mathcal{V}]$ with lex order prioritizing $\mathcal{M}$
        \STATE Declare ideal $I = \langle \mathcal{A} \rangle$
        \STATE Compute Gröbner basis $\mathcal{G} = \text{GB}(I)$

        \IF{$\mathcal{G} = \{1\}$}
            \STATE \textbf{Return} $\emptyset$ \COMMENT{System is inconsistent}
        \ENDIF

        \STATE Compute $\mathcal{G}' = \mathcal{G} \cap \mathbb{R}[\mathcal{M}]$

        \IF{$\mathcal{G}'$ is nonempty and the first element is non-monomial and contains $\leq 1$ constant}
            \STATE Select first polynomial $p \in \mathcal{G}'$
            \STATE \textbf{Return} $p$
        \ENDIF
    \ENDFOR
    \STATE $n \gets n + 1$
\ENDWHILE

\STATE \textbf{Return} $\emptyset$
\end{algorithmic}
\label{alg:projection}
\end{algorithm}

Algorithm~\ref{alg:projection} outlines our procedure for generating a consequence from a given axiom system $\mathcal{A}$ and checking its consistency. Starting with a set of variables $\mathcal{V}$, we randomly shuffle and sample a subset $\mathcal{M} \subset \mathcal{V}$ to serve as the set of measured and observed indeterminates, which are the measured variables, measured derivatives, and observed named constants combined. We require that $\mathcal{M}$ does not fully overlap with the variable set of any individual axiom in $\mathcal{A}$ to avoid trivially reproducing an axiom as a consequence. The non-measured variables are defined as $\mathcal{N} = \mathcal{V} \setminus \mathcal{M}$.

We construct a polynomial ring with a lexicographic order in which the variables in $\mathcal{M}$ appear first, and compute the Gröbner basis $\mathcal{G}$ of the ideal generated by $\mathcal{A}$ using Buchberger's algorithm~\cite{Buchberger,coxOshea}, implemented in Macaulay2~\cite{M2}. If \( \mathcal{G} = \{1\} \), this indicates inconsistency and the system is immediately discarded.

Otherwise, as stated in the elimination theorem in Section \ref{elimThm} of the main body of the paper, we compute $\mathcal{G}' = \mathcal{G}\cap \mathbb{R}[\mathcal{M}]$. If $\mathcal{G}'$ is nonempty, we consider its first polynomial as a candidate consequence. To ensure that derived consequences are informative and suitable for data generation, we apply two constraints: (i) the polynomial must not be a monomial (e.g. \( d_1 d_2 = 0 \)), and (ii) it must contain at most one observed constant. This second constraint is imposed for data scaling purposes and can be relaxed by the user (see Section~\ref{CodeAvail}).

If no valid consequence is found, we expand $\mathcal{M}$ by adding additional variables, derivatives, or constants, while still avoiding full overlaps with any axiom, and repeat the process. We continue until a valid consequence is found or until $N_{\text{max}} = 10$ attempts have been made, at which point we return $\emptyset$ and reject the axiom system.

\section{Sample Theories and Consequences} \label{app:samples}
We illustrate the theory generation process through six examples.  We specified the following sets of variables, derivatives, and named constants:
\begin{itemize}[noitemsep, topsep=0pt]
\item Variables: $Fc, Fg, W, m1, m2, d1, d2, p, \theta, \cos(\theta), \sin(\theta), e^\theta$;
\item Derivatives:  $dx1dt = \frac{dx1}{dt}, d2x1dt2 = \frac{d^2x1}{dt^2}$, 
$dx2dt = \frac{dx2}{dt}, d2x2dt2 = \frac{d^2x2}{dt^2}$; 
\item Constants: G, c.
\end{itemize}
$Fc, Fg$ denote centrifugal and gravitational forces, $w$ is a variable denoting an amount of work, $m1, m2$ denote two different masses, $d1, d2$ denote their respective distances, and $p$ denotes an amount of momentum. $dx1dt, d2x1dt2$ represent the first and second derivatives of motion associated with the first mass and $dx1dt, d2x1dt2$ represent the first and second derivatives of motion associated with the second mass. For ease of readability, we shall render the above variables and derivatives using the more conventional notation: $Fc = F_c, Fg = F_g, m1 = m_1, m2 = m_2, dx1dt = \frac{dx_1}{dt}, d2x1dt2 = \frac{d^2x_1}{dt^2}$ and so on.
$G, c$ are the gravitational constant and speed of light respectively.

The following are some sample axioms and consequences from the generated dataset. 

\begin{tcolorbox}[colback=gray!5!white, title=System 1, boxrule=0.5pt, breakable]
\begin{gather*}
\textbf{Axioms:} \\[4pt]
\frac{d^2 x_1}{dt^2}\sin{\theta} = \theta\,\frac{d^2 x_2}{dt^2}  \\[6pt]
c\,d_1\,\theta = -d_2\,\frac{dx_1}{dt}  \\[6pt]
W d_2\,\frac{dx_1}{dt} = G m_2 p \\[6pt]
d_2^2 \sin{\theta}\,\frac{dx_1}{dt} = d_1^2\,\frac{dx_2}{dt} \\[8pt]
\rule{\linewidth}{0.4pt} \\[6pt]
\textbf{Consequence (Target Polynomial):} \\[4pt]
\,\frac{dx_2}{dt} = \frac{G m_2 p d_2 \sin{\theta}}{W d_1^2}
\end{gather*}
\end{tcolorbox}

\begin{tcolorbox}[colback=gray!5!white, title=System 2, boxrule=0.5pt, breakable]
\begin{gather*}
\textbf{Axioms:} \\
\frac{dx_1}{dt}(F_g - F_c) = cF_g \\
c\frac{dx_1}{dt}F_g = F_c d_2\frac{d^2x_1}{dt^2} \\
2W^2\frac{d^2x_1}{dt^2} = d_2\frac{d^2x_1}{dt^2}F_c \\
W\frac{dx_2}{dt} = d_2F_g\frac{dx_1}{dt} \\
2d_1^2 + d_1d_2 - d_2^2 = 0 \\
\rule{\linewidth}{0.4pt} \\
\textbf{Consequence:} \\
W = \frac{cF_g^2d_2}{(F_c - F_g)\frac{dx_2}{dt}}
\end{gather*}
\end{tcolorbox}

\begin{tcolorbox}[colback=gray!5!white, title=System 3, boxrule=0.5pt, breakable]
\begin{gather*}
\textbf{Axioms:} \\[4pt]
c\,m_2 - p\,\theta = 0 \\[6pt]
m_2\,\frac{d^2 x_2}{dt^2} = F_g\,\sin{\theta}  \\[6pt]
\,\frac{d^2 x_2}{dt^2} + p^2 = d_2\,m_1 m_2 \\[6pt]
4Gc\,F_g^2 + 3G\,F_g\,p\,\frac{d^2 x_2}{dt^2}
= c^3 d_2 m_2\left(\frac{d^2 x_2}{dt^2}\right)^2  \\[6pt]
\sin{\theta}\,\theta\,\frac{dx_2}{dt} + \frac{dx_2}{dt} = c\,\theta \\[8pt]
\rule{\linewidth}{0.4pt} \\[6pt]
\textbf{Consequence:} \\[4pt]
\Bigg(
d_2\,p^{2}\theta^{2}\Big(p\theta^{2} - m_2\frac{dx_2}{dt}\Big)^{2}
\;+\;
G\,m_2^{4}\frac{dx_2}{dt}\Big(-3p\theta^{2} + m_2\frac{dx_2}{dt}(3-4\theta^{2})\Big)
\Bigg)=0
\end{gather*}
\end{tcolorbox}

\begin{tcolorbox}[colback=gray!5!white, title=System 4, boxrule=0.5pt, breakable]
\begin{gather*}
\textbf{Axioms:} \\
 2W \frac{d^2 x_2}{dt^2} + 3F_g \left(\frac{dx_2}{dt}\right)^2 = W \cos(\theta) e^{\theta} \frac{d^2 x_1}{dt^2} \\
 c =  \frac{dx_1}{dt} \cos(\theta) e^{\theta} \\
\cos(\theta) \frac{d^2 x_1}{dt^2} = \sin(\theta) \theta \frac{d^2 x_2}{dt^2}  \\
\cos^2(\theta) + \sin^2(\theta) = 1 \\
\rule{\linewidth}{0.4pt} \\
\textbf{Consequence:} \\
F_g  = \frac{\sin(\theta) W e^{\theta} \theta \frac{d^2 x_2}{dt^2} - 2W \frac{d^2 x_2}{dt^2}}{3\left(\frac{dx_2}{dt}\right)^2} 
\end{gather*}
\end{tcolorbox}

\begin{tcolorbox}[colback=gray!5!white, title=System 5, boxrule=0.5pt, breakable]
\begin{gather*}
\textbf{Axioms:} \\[4pt]
\,\frac{dx_1}{dt} + m_1 p\,\theta + 4m_1 p + 2m_2 p = c\,m_1^{2}\sin{\theta} + m_1 m_2 \\[6pt]
2W m_1\,\frac{d^2 x_2}{dt^2} = c\,F_g\,p \\[6pt]
c\,F_c\,\frac{dx_2}{dt} = W\sin{\theta}\,\theta\,\frac{d^2 x_2}{dt^2} \\[6pt]
G F_c\sin{\theta} = 2G F_c - G F_g + 3 d_2^{2}\left(\frac{d^2 x_1}{dt^2}\right)^{2} \\[6pt]
\theta\,\frac{d^2 x_2}{dt^2} = \sin{\theta}\,\frac{d^2 x_1}{dt^2} \\[8pt]
\rule{\linewidth}{0.4pt} \\[6pt]
\textbf{Consequence:} \\[4pt]
2F_c m_1\,\frac{dx_2}{dt}\,\Big(F_g + F_c\sin{\theta} - 2F_c\Big)
\;=\;
\theta\,p\,\sin{\theta}\,\Big(F_g^2\sin{\theta} + F_g F_c\sin^{2}{\theta} - 2F_g F_c\sin{\theta}\Big)
\end{gather*}
\end{tcolorbox}

\begin{tcolorbox}[colback=gray!5!white, title=System 6, boxrule=0.5pt, breakable]
\begin{gather*}
\textbf{Axioms:} \\[4pt]
m_1 m_2 \sin{\theta}\,\frac{d^2 x_2}{dt^2} + F_c m_1 + F_g m_2 = 0 \\[6pt]
p\,\frac{dx_1}{dt} + F_c d_1 = c\,m_1\,\frac{dx_2}{dt} \\[6pt]
c\,d_1 m_2\,\frac{d^2 x_2}{dt^2} + W\,\frac{dx_2}{dt} = 0 \\[6pt]
G^{2} m_1^{2}\sin{\theta} + G d_1 p\,\frac{dx_2}{dt} = G W d_1\sin{\theta} - 3G W d_1
- c^{2} d_1^{2}\sin^{2}{\theta}\left(\frac{dx_1}{dt}\right)^{2} \\[6pt]
4c\,\sin{\theta}\,\theta\,\frac{dx_2}{dt} = d_1\theta\,\frac{d^2 x_1}{dt^2} - d_1\,\frac{d^2 x_1}{dt^2} \\[6pt]
m_2\sin{\theta}\,\theta\,\frac{dx_2}{dt} + m_2\,\frac{dx_2}{dt} = p \\[8pt]
\rule{\linewidth}{0.4pt} \\[6pt]
\textbf{Consequence:} \\[4pt]
m_2\Big(F_c d_1 + p\,\frac{dx_1}{dt}\Big)\Big(1+\theta\sin{\theta}\Big) = c m_1 p
\end{gather*}
\end{tcolorbox}

\paragraph{Dataset structure and contents.}
Each configuration of variable count, derivative count, and equation count includes three generated systems, organized into folders under a consistent naming convention. For each system, we provide a symbolic description of the equations (\texttt{system.txt}), a corresponding ground-truth consequence polynomial (\texttt{consequence.txt}), and a set of replacement variants introducing minor equation modifications (\texttt{replacement\_i.txt}). Numerical datasets for both the system and its consequence are included in CSV-like files (\texttt{system.dat}, \texttt{consequence.dat}), alongside noisy variants with varying Gaussian noise levels. These files contain values for all variables, constants, and derivatives used in the equations. Each equation system is dimensionally consistent and includes metadata about units of measure, variable roles (measured, constant, derived), and the specific polynomial used as the symbolic regression target.

\section{Benchmarking Results}\label{Benchmarking Results}

The variable configurations that were tested can be found in Table~\ref{tab:models-comparison_trig} 
for systems \emph{without} trigonometric and exponential functions, and in Table~\ref{tab:models-comparison_trig} for systems \emph{with} these special functions. Performance-versus-parameter breakdowns (by number of variables, derivatives, and equations) appear in Tables~\ref{tab:var-performance_trig}, \ref{tab:deriv-performance_trig}, and \ref{tab:eq-performance_trig}.

\begin{table}[h]
    \centering
    \renewcommand{\arraystretch}{1.2}
    \caption{Performance Comparison of Models for Selected Configurations. Each Entry is of the Form (Noise = 0, Noise = $10^{-3}$, Noise = $10^{-1}$).}
    \begin{tabular}{|c|c|c|c|c|c|c|c|c|}
        \hline
        \textbf{Vars} & \textbf{Derivs} & \textbf{Eqns} & \textbf{AI Hilbert} & \textbf{AI Hilbert (replaced axioms)} & \textbf{AI Feynman} & \textbf{PySR} & \textbf{Bayesian MS} & \textbf{GPG} \\
        \hline
        6 & 2 & 4 & (\checkmark, \checkmark, x) & (\checkmark, x, x) & (\checkmark, x, x) & (\checkmark, \checkmark, x) & (x, x, x) & (\checkmark, x, x) \\
        6 & 2 & 5 & (\checkmark, \checkmark, \checkmark) & (\checkmark, \checkmark, \checkmark) & (\checkmark, \checkmark, x) & (\checkmark, \checkmark, x) & (\checkmark, \checkmark, \checkmark) & (x, x, x) \\
        6 & 2 & 6 & (\checkmark, \checkmark, \checkmark) & (\checkmark, \checkmark, \checkmark) & (\checkmark, x, x) & (\checkmark, \checkmark, x) & (x, x, x) & (\checkmark, \checkmark, x) \\
        6 & 3 & 4 & (\checkmark, \checkmark, x) & (x, x, x) & (\checkmark, x, x) & (\checkmark, x, x) & (x, x, x) & (x, x, x) \\
        6 & 3 & 5 & (\checkmark, \checkmark, \checkmark) & (\checkmark, \checkmark, x) & (\checkmark, \checkmark, \checkmark) & (\checkmark, \checkmark, \checkmark) & (\checkmark, \checkmark, \checkmark) & (\checkmark, \checkmark, \checkmark) \\
        6 & 3 & 6 & (\checkmark, \checkmark, x) & (\checkmark, x, x) & (\checkmark, x, x) & (x, x, x) & (x, x, x) & (\checkmark, \checkmark, x) \\
        6 & 4 & 4 & (\checkmark, \checkmark, x) & (\checkmark, \checkmark, x) & (x, x, x) & (x, x, x) & (x, x, x) & (\checkmark, x, x) \\
        6 & 4 & 5 & (\checkmark, x, x) & (x, x, x) & (x, x, x) & (x, x, x) & (x, x, x) & (x, x, x) \\
        6 & 4 & 6 & (\checkmark, \checkmark, \checkmark) & (x, x, x) & (\checkmark, x, x) & (\checkmark, \checkmark, x) & (x, x, x) & (x, x, x) \\
        7 & 2 & 4 & (x, x, x) & (x, x, x) & (x, x, x) & (x, x, x) & (x, x, x) & (x, x, x) \\
        7 & 2 & 5 & (\checkmark, \checkmark, x) & (\checkmark, x, x) & (\checkmark, x, x) & (x, x, x) & (\checkmark, \checkmark, x) & (\checkmark, \checkmark, x) \\
        7 & 2 & 6 & (x, x, x) & (x, x, x) & (x, x, x) & (x, x, x) & (x, x, x) & (\checkmark, x, x) \\
        7 & 3 & 4 & (\checkmark, x, x) & (x, x, x) & (\checkmark, x, x) & (\checkmark, \checkmark, x) & (x, x, x) & (x, x, x) \\
        7 & 3 & 5 & (\checkmark, \checkmark, x) & (x, x, x) & (x, x, x) & (\checkmark, x, x) & (\checkmark, \checkmark, x) & (\checkmark, x, x) \\
        7 & 3 & 6 & (\checkmark, \checkmark, \checkmark) & (\checkmark, \checkmark, x) & (\checkmark, \checkmark, x) & (\checkmark, \checkmark, x) & (\checkmark, \checkmark, \checkmark) & (\checkmark, x, x) \\
        7 & 4 & 4 & (x, x, x) & (x, x, x) & (\checkmark, x, x) & (\checkmark, \checkmark, x) & (x, x, x) & (x, x, x) \\
        7 & 4 & 5 & (x, x, x) & (x, x, x) & (x, x, x) & (\checkmark, x, x) & (x, x, x) & (x, x, x) \\
        7 & 4 & 6 & (\checkmark, x, x) & (x, x, x) & (x, x, x) & (x, x, x) & (x, x, x) & (x, x, x) \\
        8 & 2 & 4 & (\checkmark, \checkmark, x) & (x, x, x) & (\checkmark, x, x) & (\checkmark, x, x) & (\checkmark, x, x) & (\checkmark, x, x) \\
        8 & 2 & 5 & (\checkmark, \checkmark, x) & (x, x, x) & (\checkmark, x, x) & (x, x, x) & (x, x, x) & (\checkmark, x, x) \\
        8 & 2 & 6 & (x, x, x) & (x, x, x) & (\checkmark, x, x) & (\checkmark, x, x) & (x, x, x) & (x, x, x) \\
        8 & 3 & 4 & (x, x, x) & (x, x, x) & (x, x, x) & (x, x, x) & (x, x, x) & (x, x, x) \\
        8 & 3 & 5 & (x, x, x) & (x, x, x) & (x, x, x) & (x, x, x) & (x, x, x) & (x, x, x) \\
        8 & 3 & 6 & (\checkmark, x, x) & (x, x, x) & (x, x, x) & (x, x, x) & (\checkmark, x, x) & (x, x, x) \\
        8 & 4 & 4 & (x, x, x) & (x, x, x) & (x, x, x) & (x, x, x) & (x, x, x) & (x, x, x) \\
        8 & 4 & 5 & (\checkmark, \checkmark, \checkmark) & (\checkmark, \checkmark, x) & (x, x, x) & (\checkmark, x, x) & (x, x, x) & (x, x, x) \\
        8 & 4 & 6 & (\checkmark, \checkmark, \checkmark) & (\checkmark, \checkmark, \checkmark) & (\checkmark, x, x) & (\checkmark, \checkmark, x) & (\checkmark, x, x) & (\checkmark, \checkmark, x) \\
        9 & 2 & 4 & (x, x, x) & (x, x, x) & (x, x, x) & (x, x, x) & (x, x, x) & (x, x, x) \\
        9 & 2 & 5 & (\checkmark, \checkmark, x) & (\checkmark, x, x) & (\checkmark, x, x) & (x, x, x) & (\checkmark, x, x) & (x, x, x) \\
        9 & 2 & 6 & (x, x, x) & (x, x, x) & (x, x, x) & (x, x, x) & (x, x, x) & (x, x, x) \\
        9 & 3 & 4 & (\checkmark, x, x) & (\checkmark, x, x) & (x, x, x) & (x, x, x) & (x, x, x) & (\checkmark, x, x) \\
        9 & 3 & 5 & (x, x, x) & (x, x, x) & (x, x, x) & (x, x, x) & (x, x, x) & (x, x, x) \\
        9 & 3 & 6 & (\checkmark, \checkmark, \checkmark) & (x, x, x) & (x, x, x) & (x, x, x) & (x, x, x) & (\checkmark, x, x) \\
        9 & 4 & 4 & (x, x, x) & (x, x, x) & (x, x, x) & (x, x, x) & (x, x, x) & (x, x, x) \\
        9 & 4 & 5 & (\checkmark, x, x) & (x, x, x) & (x, x, x) & (x, x, x) & (\checkmark, x, x) & (x, x, x) \\
        9 & 4 & 6 & (x, x, x) & (x, x, x) & (x, x, x) & (x, x, x) & (x, x, x) & (x, x, x) \\
        \hline
    \end{tabular}
    \label{tab:models-comparison_trig}
\end{table}

\begin{table}[h]
    \centering
    \renewcommand{\arraystretch}{1.2}
    \caption{Performance by Number of Variables. Entries show successes without noise, with $\epsilon=10^{-3}$, with $\epsilon=10^{-1}$ out of total configurations for each column.}
    \begin{tabular}{|c|c|c|c|c|}
        \hline
        \textbf{Method} & \textbf{6 Variables} & \textbf{7 Variables} & \textbf{8 Variables} & \textbf{9 Variables} \\
        \hline
        AI Hilbert & (9/9, 8/9, 4/9) & (5/9, 3/9, 1/9) & (5/9, 4/9, 2/9) & (4/9, 2/9, 1/9) \\
        AI Hilbert w/ replaced axioms & (6/9, 4/9, 2/9) & (2/9, 1/9, 0/9) & (2/9, 2/9, 1/9) & (2/9, 0/9, 0/9) \\
        AI Feynman & (7/9, 2/9, 1/9) & (4/9, 1/9, 0/9) & (4/9, 0/9, 0/9) & (1/9, 0/9, 0/9) \\
        PySR & (6/9, 5/9, 1/9) & (5/9, 3/9, 0/9) & (4/9, 1/9, 0/9) & (0/9, 0/9, 0/9) \\
        Bayesian MS & (2/9, 2/9, 2/9) & (3/9, 3/9, 1/9) & (3/9, 0/9, 0/9) & (2/9, 0/9, 0/9) \\
        GPG & (5/9, 3/9, 1/9) & (4/9, 1/9, 0/9) & (3/9, 1/9, 0/9) & (2/9, 0/9, 0/9) \\
        \hline
    \end{tabular}
    \label{tab:var-performance_trig}
\end{table}

\begin{table}[h]
    \centering
    \renewcommand{\arraystretch}{1.2}
    \caption{Performance by Number of Derivatives. Entries show successes without noise, with $\epsilon=10^{-3}$, with $\epsilon=10^{-1}$ out of total configurations for each column.}
    \begin{tabular}{|c|c|c|c|}
        \hline
        \textbf{Method} & \textbf{2 Derivatives} & \textbf{3 Derivatives} & \textbf{4 Derivatives} \\
        \hline
        AI Hilbert & (7/12, 7/12, 2/12) & (9/12, 6/12, 3/12) & (7/12, 4/12, 3/12) \\
        AI Hilbert w/ replaced axioms & (5/12, 2/12, 2/12) & (4/12, 2/12, 0/12) & (3/12, 3/12, 1/12) \\
        AI Feynman & (8/12, 1/12, 0/12) & (5/12, 2/12, 1/12) & (3/12, 0/12, 0/12) \\
        PySR & (5/12, 3/12, 0/12) & (5/12, 3/12, 1/12) & (5/12, 3/12, 0/12) \\
        Bayesian MS & (4/12, 2/12, 1/12) & (4/12, 3/12, 2/12) & (2/12, 0/12, 0/12) \\
        GPG & (6/12, 2/12, 0/12) & (6/12, 2/12, 1/12) & (2/12, 1/12, 0/12) \\
        \hline
    \end{tabular}
    \label{tab:deriv-performance_trig}
\end{table}

\begin{table}[h]
    \centering
    \renewcommand{\arraystretch}{1.2}
    \caption{Performance by Number of Equations. Entries show successes without noise, with $\epsilon=10^{-3}$, with $\epsilon=10^{-1}$ out of total configurations for each column.}
    \begin{tabular}{|c|c|c|c|}
        \hline
        \textbf{Method} & \textbf{4 Equations} & \textbf{5 Equations} & \textbf{6 Equations}  \\
        \hline
        AI Hilbert & (6/12, 4/12, 0/12) & (9/12, 7/12, 3/12) & (8/12, 6/12, 5/12) \\
        AI Hilbert (replaced) & (3/12, 1/12, 0/12) & (5/12, 3/12, 1/12) & (4/12, 3/12, 2/12) \\
        AI Feynman & (5/12, 0/12, 0/12) & (5/12, 2/12, 1/12) & (6/12, 1/12, 0/12) \\
        PySR & (5/12, 3/12, 0/12) & (5/12, 2/12, 1/12) & (5/12, 4/12, 0/12) \\
        Bayesian MS & (1/12, 0/12, 0/12) & (6/12, 4/12, 2/12) & (3/12, 1/12, 1/12) \\
        GPG & (4/12, 0/12, 0/12) & (4/12, 2/12, 1/12) & (6/12, 3/12, 0/12) \\
        \hline
    \end{tabular}
    \label{tab:eq-performance_trig}
\end{table}

For the trig/exponential set (Table~\ref{tab:models-comparison_trig}), \textbf{AI Hilbert} solves $23/36$ ($\approx\!64\%$) cases at $\epsilon=0$, $17/36$ ($\approx\!47\%$) at $\epsilon=10^{-3}$, and $8/36$ ($\approx\!22\%)$ at $\epsilon=10^{-1}$. With incorrect (replaced) axiom systems, performance drops to $12/36$ ($\approx\!33\%$), $7/36$ ($\approx\!19\%$), and $3/36$ ($\approx\!8\%)$, respectively. \textbf{PySR} correspondingly attains $15/36$ ($\approx\!42\%$), $9/36$ ($\approx\!25\%$), and $1/36$ ($\approx\!3\%)$, while \textbf{AI Feynman} attains $16/36$ ($\approx\!44\%$), $3/36$ ($\approx\!8\%$), and $1/36$ ($\approx\!3\%)$. The newly benchmarked methods show varying performance: \textbf{Bayesian Machine Scientist} correspondingly achieves $10/36$ ($\approx\!28\%$), $5/36$ ($\approx\!14\%$), and $3/36$ ($\approx\!8\%)$, while \textbf{GPG} achieves $14/36$ ($\approx\!39\%$), $5/36$ ($\approx\!14\%$), and $1/36$ ($\approx\!3\%)$.

We see in Tables~\ref{tab:var-performance_trig}, \ref{tab:deriv-performance_trig}, and \ref{tab:eq-performance_trig} that the most significant drop in performance for PySR and AI Hilbert comes with increasing the number of variables in the axiom system. PySR and AI Feynman are agnostic to the number of equations in the background theory and do not take them into account during search. AI Hilbert does account for the number of equations but is largely unaffected by a small change from $4$ to $5$. Performance is not sharply affected by the number of derivatives, though the proportion of correct solutions declines as the number of variables and derivatives increases.  Bayesian Machine Scientist shows relatively stable performance across different numbers of derivatives and equations, with notable resilience to noise compared to AI Feynman and PySR. GPG demonstrates performance comparable to PySR in the noiseless case but experiences similar degradation with increasing noise levels.

We note some factors that might affect the performance of the models. The first is that these systems were run on different hardware. AI Hilbert and PySR were run on an Apple Macbook Pro M2 system with 8 cores and 16GB RAM on 4 threads, while AI Feynman was run on a high-performance computing node with Intel Xeon Platinum 8260 processors with 192GB RAM on 48 threads. Bayesian Machine Scientist and GPG were run on a server comprising two Intel Xeon (Cascadelake) processors with 256GB RAM on 64 threads.

The second factor is the runtime. We imposed a 2-hour wall-clock timeout constraint on PySR, AI Hilbert, Bayesian Machine Scientist, and GPG. For each attempted solution, AI Hilbert averaged a CPU time of approximately 30 minutes, PySR 1 hour and 15 minutes, and AI Feynman 15 hours. Given that the models were run on different hardware, the CPU time to wall-clock time upper limit (only imposed on PySR and AI Hilbert) is not one-to-one across the models. Therefore, while AI Feynman averaged a wall-clock time of 1 hour and 30 minutes per attempted solution, relatively closer to the wall-clock times of PySR and AI Hilbert~--~1 hour and 40 minutes respectively~--~the CPU time was drastically higher due to the compute resources dedicated to it. The average wall-clock time for Bayesian Machine Scientist was 17 minutes and for GPG was 2 seconds. When the time constraint is exceeded, for AI Hilbert, we count these as failure cases since these models do not provide intermediate results. Since PySR does provide intermediate results, upon exceeding the time constraint, we consider the best five candidate equations available and compare to the true solution. In our experimentation, when PySR exceeded the time constraint, the available candidates did not correspond to the true solution, and hence were counted as failure cases. Since no time constraint was used for AI Feynman, the failure cases were either the model failing to converge to a solution or converging to the incorrect solution. All runs with Bayesian Machine Scientist and GPG finished before the time limit was reached. There were a few cases where Bayesian Machine Scientist stopped with an error and did not return any expression.

\textcolor{black}{The third factor is that each model was run with the default setup publicly available in the respective github repositories associated with the papers except for a small change done to AI Feynman, which consisted in commenting out lines in the code calling functions of the type \texttt{get\_X} (where \texttt{X} is replaced by \texttt{sin}, \texttt{cos}, etc.) suggested by one of the authors of AI Feynman to avoid the generation of solutions with trigonometric and other special functions in the cases from Table \ref{tab:models-comparison_trig} 
where no special functions were used in the systems. We only made that change to the off-the-shelf AI Feynman architecture and only made hyperparameter changes to the available models.} 
For AI Feynman, we set the \textit{BF try time}~--~time limit for each brute force call~--~to 30 seconds, the maximum degree of the polynomial fit routing to 3, the number of epochs for training to 100, and we used the default list of operations in the brute force search (Add (+), Multiply (*), Subtract (-), Divide (/), Increment (>), Decrement (<), Invert (I), sqrt (R)). For AI Hilbert, we configured the model with the Gurobi optimizer (version 9.5) using default tolerance settings and sparsity constraints. For PySR, we used a population size of 400 candidate expressions evolved over 750 generations, with expression complexity limited to 100 nodes and parsimony coefficient of 0.01. For Bayesian Machine Scientist, we used the prior hyperparameters with the number of parameters equal to the number of variables plus one. For GPG, we added the operation sqrt to the default list of operations. These hardware and hyperparameter constraints were chosen to yield comparable results. But it is surely the case that further fine tuning of the neural models as well as of the optimization-based models, along with running on uniform hardware with longer run-times, would have affected the results.

\section{Sensitivity Analysis: Robustness of SR Systems to Different Types of Noise}
\label{sec:sensitivity-analysis}

\subsection{Methodology} 
\label{sec:sensitivity-analysis-methodology}

To evaluate the robustness of the various SR systems with respect to different \emph{types} of noisy data (in addition to different magnitudes of noise), we conducted a sensitivity analysis using exponential and log-normal distributions for the noise production models in addition to the more traditional Gaussian noise distribution. 
We started by using two well-established physical theories with known axiom systems: 
\textbf{Newtonian Gravity} (the theory used to derive Kepler's Third law of Planetary Motion in \cite{ai-descartes23}) and \textbf{Lorentz-Invariant Motion} (the theory used to derive Einstein's Relativistic Time Dilation Law in \cite{ai-hilbert24}).
For each of these two theories, we replaced one axiom with a randomly generated, dimensionally consistent axiom to create an alternate ground truth system. Following the methodology described earlier in this paper, we generated consequences from the ground truth system using dimensional analysis and Gr\"{o}bner basis-based algebraic manipulation. We then generated exact data associated with this consequence using previously described methods, and also superimposed noise of the three distributional types, each at a ``low noise'' and ``high noise'' level. (We describe shortly what is meant respectively by ``low'' and ``high'' noise levels for each model.) Lastly, we benchmarked the same set of SR systems we have used for the rest of this study, namely \textbf{AI Hilbert}, \textbf{AI Hilbert (with replaced/incorrect axioms)}, \textbf{AI Feynman}, \textbf{PySR}, \textbf{Bayesian Machine Scientist}, and \textbf{GPG} to see if they could successfully recover the consequence in each scenario. \textbf{AI Hilbert (with replaced/incorrect axioms)} is the same as \textbf{AI Hilbert} but where one of its assumed axioms is incorrect. 

The purpose of testing three distinct noise distributions is to assess robustness across different types of measurement uncertainties that may arise in physical systems. \textbf{Gaussian Noise} represents the standard normal distribution commonly assumed in classical error analysis.
For comparison, we consider \textbf{Exponential Noise} and \textbf{Log-Normal Noise} models to evaluate the robustness of the various SR systems under heavier distributional tails for the different types of noisy data.
In particular, our choice of the exponential distribution allows us to consider a noise distribution with a heavier tail than the commonly used Gaussian distribution.
Similarly, our choice of the log-normal distribution allows us to consider a noise distribution with a heavier tail than the exponential distribution.
We note that the Gaussian distribution is symmetric and provides both positive and negative values, while both the exponential and log-normal distributions only provide positive values.
To address this and provide comparable noise models, we flip a fair coin to determine whether the exponential or log-normal noise will be positive or negative, and then accordingly apply the sampled noise from the corresponding distribution.
Since the noise will be positive with probability $0.5$ and negative with probability $0.5$, this results in a symmetric noise model under both the exponential and log-normal distributions similar in this respect to the Gaussian distribution.

For each noise model and each theory, we evaluated performance across multiple noise levels to understand how the methods degrade as measurement uncertainty increases. The analysis therefore provides both a cross-method comparison and an assessment of generalization to realistic noise profiles beyond standard Gaussian assumptions. 
The non-zero noise levels for the Gaussian distribution are $10^{-3}$ (low noise) and $10^{-1}$ (high noise) as listed in Table~\ref{tab:summary-benchmark}.
For comparison, we set the parameter $\lambda$ of the exponential distribution to be $\lambda = \sqrt{1/\sigma^2}$ where $\sigma^2$ is either the low-noise variance or high-noise variance of the Gaussian distribution.
Similarly, we set the parameters $\mu$ and $\sigma$ of the log-normal distribution such that $\mu=0$ and $\sigma$ is chosen so that the variance of the log-normal distribution,  $(e^{\sigma^2} - 1) e^{\sigma^2}$, is twice either the low-noise variance or high-noise variance of the Gaussian distribution.
Hence, we have the ``low noise'' and ``high noise'' levels of all three noise models scaled with respect to the Gaussian noise. 


\subsection{Results}

\subsubsection{Newtonian Gravity}

Performance results for the different SR system in the case of Newtonian Gravity under different noise types and magnitudes are presented in Table~\ref{tab:sensitivity-kepler}. Each entry shows the number of successes out of 5 attempted system instances for that noise type and level.

\begin{table}[h]
    \centering
    \renewcommand{\arraystretch}{1.2}
    \caption{Sensitivity Analysis: Newtonian Gravity. Each Entry is of the Form (Gaussian, Log-Normal, Exponential). Entries show successes with low noise and high noise levels of $\epsilon=10^{-3}$ and $\epsilon=10^{-1}$ for the Gaussian noise model, and with the corresponding exponential and log-normal noise models scaled accordingly as described in Appendix~\ref{sec:sensitivity-analysis-methodology}.
    }
    \begin{tabular}{|c|c|c|c|}
        \hline
        \textbf{Method} & \textbf{No Noise} & \textbf{Low Noise} & \textbf{High Noise} \\
        \hline
        AI Hilbert & 4/5 & (3/5, 3/5, 4/5) & (2/5, 1/5, 0/5) \\
        AI Hilbert (replaced axioms) & 2/5 & (1/5, 0/5, 1/5) & (0/5, 0/5, 0/5) \\
        AI Feynman & 2/5 & (1/5, 0/5, 1/5) & (0/5, 0/5, 0/5) \\
        PySR & 3/5 & (1/5, 1/5, 2/5) & (0/5, 1/5, 1/5) \\
        Bayesian Machine Scientist & 3/5 & (2/5, 3/5, 1/5) & (2/5, 0/5, 0/5) \\
        GPG & 3/5 & (2/5, 1/5, 2/5) & (2/5, 1/5, 2/5) \\
        \hline
    \end{tabular}
    \label{tab:sensitivity-kepler}
\end{table}

\subsubsection{Lorentz-Invariant Motion}

Performance results for the different SR systems in the case of Lorentz-Invariant Motion under different noise types and magnitudes are presented in Table~\ref{tab:sensitivity-einstein}. Each entry shows the number of successes out of 5 attempted system instances for that noise type and level.

\begin{table}[h]
    \centering
    \renewcommand{\arraystretch}{1.2}
    \caption{Sensitivity Analysis: Lorentz-Invariant Motion. Each Entry is of the Form (Gaussian, Log-Normal, Exponential). Entries show successes with low noise and high noise levels of $\epsilon=10^{-3}$ and $\epsilon=10^{-1}$ for the Gaussian noise model, and with the corresponding exponential and log-normal noise models scaled accordingly as described in Appendix~\ref{sec:sensitivity-analysis-methodology}.
    }
    \begin{tabular}{|c|c|c|c|}
        \hline
        \textbf{Method} & \textbf{No Noise} & \textbf{Low Noise} & \textbf{High Noise} \\
        \hline
        AI Hilbert & 3/5 & (2/5, 1/5, 2/5) & (1/5, 1/5, 0/5) \\
        AI Hilbert (replaced axioms) & 2/5 & (1/5, 2/5, 0/5) & (1/5, 0/5, 0/5) \\
        AI Feynman & 3/5 & (1/5, 1/5, 0/5) & (0/5, 0/5, 0/5) \\
        PySR & 2/5 & (0/5, 0/5, 1/5) & (0/5, 0/5, 0/5) \\
        Bayesian Machine Scientist & 2/5 & (2/5, 2/5, 1/5) & (1/5, 1/5, 1/5) \\
        GPG & 3/5 & (1/5, 1/5, 1/5) & (1/5, 1/5, 1/5) \\
        \hline
    \end{tabular}
    \label{tab:sensitivity-einstein}
\end{table}

\subsection{Discussion}

Interestingly, in the low noise level environment, most of the models do not perform very much worse (and in some cases perform marginally better) when the noise production model veers away from the standard Gaussian distribution. We note, however, that in high noise level environments, both AI Hilbert and the Bayesian Machine Scientist performed notably worse, for the use case of Newtonian Gravity, when noise production was non-Gaussian compared with when the noise production was Gaussian.

\section{Computational Resources Used to Generate Our Axiom Systems and Data}\label{computation_resources}

Axiom system and data generation were run on an Apple MacBook Pro with an M2 chip (8-core CPU) and 16GB of unified memory on a single thread, generating 108 complete systems and associated data. Table \ref{tab:performance} summarizes the performance characteristics.

\begin{table}[htbp] \label{tab:perf-metrics}
    \centering
    \caption{Dataset Generation Performance Characteristics.}
    \label{tab:performance}
    \begin{tabular}{lr}
        \hline
        \textbf{Metric} & \textbf{Value} \\
        \hline
        Total CPU time & 
        22.15 hrs \\
        Peak Memory Usage & 137.03 MB \\
        \hline
        \textbf{System Component} & \textbf{CPU Time Taken} \\
        \hline
        Axiom System Generation & 
        9.44 hrs\\
        Replacement Axiom Generation & 
        12.63 hrs\\
        Consequence Generation & 
        2.10 min\\
        Data Generation for  Consequences & 
        8.27 min\\
       Biased Data Generation for Systems & 
       4.51 min\\
        \hline
    \end{tabular}
\end{table}

The performance metrics demonstrate efficient memory utilization (under 140 MB at peak usage) with computationally intensive phases, particularly replacement axiom generation taking the longest. We generate five replacement axioms per system, which for the systems generated is equal or almost equal to the initial number of axioms in each system. The generation of a replacement axiom is a somewhat more constrained problem than generating a random axiom, since the replacement axiom must include any variables, derivatives, and constants that are solely used by the axiom it replaces. Moreover, the axiom must not be a consequence of the prior axiom system.  The total CPU time for generating all 216 systems was approximately 22.15 hours. 

\textbf{Licenses}

Our code will be publicly available under an MIT License on github. This work uses third-party software:
\begin{itemize} 
  \item \textbf{SymPy} (BSD 3-Clause). License: \url{https://github.com/sympy/sympy/blob/master/LICENSE}.
  \item \textbf{Macaulay2} (GNU GPL, v2 or later / v3). License: \url{https://macaulay2.com/Downloads/Copyright/}.
\end{itemize}

\end{document}